\documentclass[a4paper,11pt]{book}
\usepackage[utf8]{inputenc}
\usepackage{fullpage}
\usepackage{fp}
\usepackage{hyperref}
\usepackage{color}
\usepackage{url}
\usepackage{graphicx}
\usepackage{fancyhdr}
\usepackage{framed}
\usepackage{lastpage}
\usepackage{listings}
\usepackage{alltt}
\usepackage{amssymb}

\advance\topmargin by -3mm

\definecolor{mancoosi@lightblue}{RGB}{1,94,140}
\definecolor{mancoosi@green}{RGB}{157,199,218}
\definecolor{mancoosi@red}{RGB}{189,16,11}

\long\def\ignore#1{\relax}

\newcommand{\EMAIL}[1]{\href{mailto:#1}{\texttt{#1}}}

\newcommand{\HOLE}[1]{\fbox{#1}}
\newcommand{\NONNORM}{\hfill{\normalsize (non-normative)}}

\newcommand{\mancoosi}{Mancoosi}

\newcommand{\NT}[1]{\ensuremath{\mbox{\small\textsc{#1}}}} 

\newenvironment{outline}{\begin{list}
 {-}
 {\setlength{\itemsep}{0mm} \setlength{\parsep}{0mm}}}
 {\end{list}}

\lstdefinestyle{XML}{language=XML,basicstyle=\small\ttfamily}
\newcommand{\PROPERTYBOX}[5]{%
  \begin{center}
    \begin{tabular}{lp{0.75\textwidth}}
      \hline
      \textsc{Name} & \texttt{#1} \\
      \textsc{Type} & \texttt{#2} \\
      \textsc{Optionality} & #3 \\
      \if#4\else \textsc{Default} & #4 \\\fi
      \textsc{Description} & #5 \\
      \hline
    \end{tabular}
\end{center}}

\newcommand{\TYPEBOX}[5]{%
  \begin{center}
    \begin{tabular}{lp{0.75\textwidth}}
      \hline
      \textsc{Name} & \texttt{#1} \\
      \textsc{Description} & #2 \\
      \textsc{Value space} & #3 \\
      \textsc{Lexical space} & #4 \\
      \textsc{Parsing} & #5 \\
      \hline
    \end{tabular}
\end{center}}

\newcommand{\LEX}[1]{\ensuremath{\mbox{\tt"#1"}}}
\newcommand{\XSTYPE}[1]{\ensuremath{\mathtt{#1}^\mathrm{\scriptscriptstyle
      XML~Schema}}}
\newcommand{\SUBTY}{\ensuremath{\mathtt{<:}}}

\newtheorem{definition}{Definition}
\newtheorem{lemma}{Lemma}

\newcommand{\DomPackages}{\ensuremath{\mathcal{V}}(\texttt{pkgname})}
\newcommand{\DomVersions}{\ensuremath{\mathcal{V}}(\texttt{posint})}
\newcommand{\DomKeep}{\textsc{Keepvalues}}
\newcommand{\DomBool}{\ensuremath{\mathcal{V}}(\texttt{bool})}
\newcommand{\DomForm}{\ensuremath{\mathcal{V}}(\texttt{vpkgformula})}
\newcommand{\DomCList}{\ensuremath{\mathcal{V}}(\texttt{vpkglist})}
\newcommand{\DomEList}{\ensuremath{\mathcal{V}}(\texttt{vepkglist})}
\newcommand{\DomConstraints}{\textsc{Constraints}}
\newcommand{\dom}{\textrm{Dom}}
\newcommand{\ValKeep}[1]{#1.\texttt{keep}}
\newcommand{\ValDepends}[1]{#1.\texttt{depends}}
\newcommand{\ValProvides}[1]{#1.\texttt{provides}}
\newcommand{\ValConflicts}[1]{#1.\texttt{conflicts}}
\newcommand{\ValInstalled}[1]{#1.\texttt{installed}}
\newcommand{\ValCost}[1]{#1.\texttt{cost}}
\newcommand{\PackageDescriptions}{\textsc{Descr}}

\newcommand{\ReqSem}[1]{\stackrel{#1}\curvearrowright}

\begin{document}
\begin{titlepage}
%
%
\begin{minipage}[p]{7cm}
\begin{large}
\textcolor{mancoosi@lightblue}{
\begin{tabular}{l}
Specific Targeted Research Project\\
Contract no.214898\\
Seventh Framework Programme: FP7-ICT-2007-1\\ 
\end{tabular}
}
\end{large}
\end{minipage}
\hfill
\includegraphics*[width=4cm]{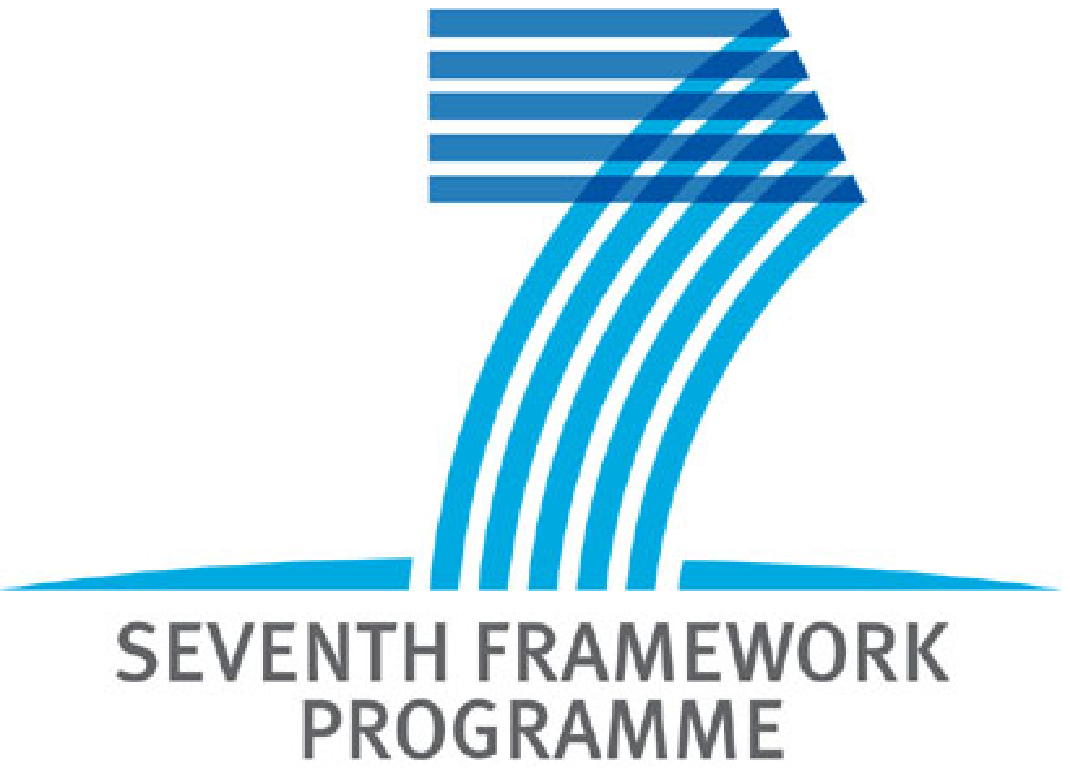} 

%
%

\includegraphics*[width=5cm]{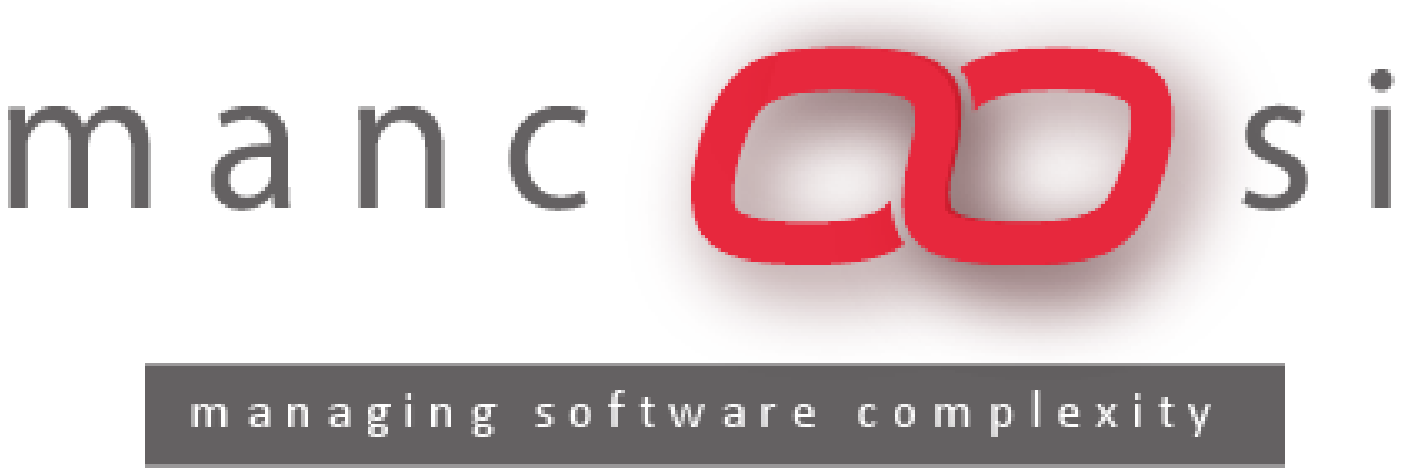} \hfill \textcolor{mancoosi@green}{\LARGE\bf Deliverable D5.1}

\vfill

\begin{center}

%
%
{\bf \Large MANCOOSI} \\[3em]

{\bf \Large Managing the Complexity of the Open Source Infrastructure}\\
\end{center}

\vfill 
%
%
\begin{framed}
\centering{\textcolor{mancoosi@lightblue}{\Huge\bf Description of the CUDF Format}}
\end{framed}

\vfill

\begin{center}
\large \today
\end{center}

\vfill

\begin{flushright}
\begin{tabular}{|l|l|}
  \hline
  \bf{Project acronym}        & MANCOOSI\\\hline
  \bf{Project full title}     & Managing the Complexity of the Open Source Infrastructure \\\hline
  \bf{Project number}         & 214898 \\\hline
  \bf{Authors list}           & Ralf Treinen
                                \verb+<+\EMAIL{Ralf.Treinen@pps.jussieu.fr}\verb+>+ \\
                              & Stefano Zacchiroli
                                \verb+<+\EMAIL{zack@pps.jussieu.fr}\verb+>+ \\\hline
  \bf{Internal review}        & Pietro Abate, Sophie Cousin,
                                Olivier Lhomme, Claude Michel,\\
                              & Jean Charles Régin, Michel Rueher 
 \\\hline
  \bf{Workpackage number}     & WP5 \\\hline
  \bf{Deliverable number}     & 1 \\\hline
  \bf{Document type}          & Deliverable \\\hline
  \bf{Version}                & $1$ \\\hline
  \bf{Due date}               & 01/11/2008\\\hline
  \bf{Actual submission date} & 01/11/2008\\\hline
  \bf{Distribution}           & Public\\\hline
  \bf{Project coordinator}    & Roberto Di Cosmo
                                \verb+<+\EMAIL{roberto@dicosmo.org}\verb+>+\\\hline
\end{tabular}
\end{flushright}

\vfill

{\bf\large Web site: \url{www.mancoosi.org}} \hfill {\bf\large Blog: \url{blog.mancoosi.org}}

\end{titlepage}

\section*{Abstract}
This document contains several related specifications, together
they describe the document formats related to the solver competition
which will be organized by Mancoosi.

In particular, this document describes:
\begin{description}
\item[DUDF] (Distribution Upgradeability Description Format), the
  document format to be used to submit upgrade problem instances from
  user machines to a (distribution-specific) database of upgrade
  problems;
\item[CUDF] (Common Upgradeability Description Format), the document
  format used to encode upgrade problems, abstracting over
  distribution-specific details. Solvers taking part in the
  competition will be fed with input in CUDF format.
\end{description}

\vfill
\section*{Conformance}

\begin{quotation}
  The key words ``MUST'', ``MUST NOT'', ``REQUIRED'', ``SHALL'',
  ``SHALL NOT'', ``SHOULD'', ``SHOULD NOT'', ``RECOMMENDED'', ``MAY'',
  and ``OPTIONAL'' in this document are to be interpreted as described
  in RFC 2119~\cite{rfc2119}.
\end{quotation}

\vfill

\tableofcontents
\listoffigures
\chapter{Introduction}
\label{chap:intro}

The aim of work package 5 (WP5) of the \mancoosi{} project is to
organize a solver competition to attract the attention of researchers
and practitioners to the upgrade problem as it is faced by users of
F/OSS distributions~\cite{mancoosi-wp1d1}. The competition will be run
by executing solvers submitted by the participants on upgrade problem
descriptions (or ``problems'', for short) stored in upgradeability
problem data bases (UPDBs). A substantial part of the problems forming
UPDBs, if not all of them, will be real problems harvested on user
machines; users will be given tools to submit on a voluntary basis
problems to help \mancoosi{} assemble UPDBs.

In such a scenario, problem descriptions need to be saved on
filesystems (for long term storage) and transmitted over the network
(to let them flow from user machines to UPDBs). This document gives
the specifications of document formats used to represent problem
instances in the various stages of their lives.

\section{Two different upgrade description formats}

Upgrade description formats serve at least two different purposes:

\begin{description}

\item[Problem submission] problems will be created on distant user
  machines and need to flow to more centralized UPDBs. Both the user
  machine itself and the network connection may have only limited
  resources.

\item[Problem description] problems will be stored by \mancoosi{} to
  form a corpus of problems on which the solvers taking part in the
  competition will be run.

\end{description}

In the \mancoosi{} Description of Work we announced the definition of
a so-called \emph{Common Upgradeability Description Format},
abbreviated CUDF, that would serve these two purposes. It turned out
that having one single format for both purposes is not practical since
both purposes come with contradicting constraints: problem submissions
should take as few resources as possible on a user's machine, and
they may contain references that are meaningful only in the context of
a particular distribution. On the other hand, problem descriptions as used for the
competition are not subject to strong resource
limitations but must be self-contained and must have a formally defined
semantics that is independent from any particular distribution.

As a consequence, we decided to define two different formats, one for
each of the main purpose:

\begin{description}

\item[DUDF (Distribution Upgradeability Description Format)] This is
  the format used to submit a single problem from user machines to a
  UPDB. DUDF is specialized for the purpose of problem submission.

  DUDF instances (or ``DUDFs'' for short) need to be as compact as
  possible in order to avoid inhibiting submissions due to excessive
  bandwidth requirements. To this end, the DUDF specification exploits
  distribution-specific information, such as the knowledge of where
  distribution-wide metadata are stored and where metadata about old
  packages can be retrieved from mirrors that may or may not be
  specific to \mancoosi.

  Since a DUDF is by its very nature distribution dependent there
  cannot be a a single complete DUDF specification.  We rather present
  in Chapter~\ref{chap:dudf} a generic specification of DUDF
  documents, the \emph{DUDF skeleton}, which has to be instantiated to
  a full specification by all participating distributions.  Documents
  to be published separately, one per distribution, will describe how
  the general scheme is instantiated by the various distributions.

  All in all we have a \emph{family of DUDF specification instances}:
  Debian-DUDF, RPM-DUDF, etc.; one for each possible way of
  filling the holes of the generic DUDF specification. How many
  instances should be part of the DUDF family? We recommend to have
  one instance for each distribution taking part in the
  competition. While different distributions may share a common
  packaging format, they may also allow for different means of compact
  representations, for example due to the different availability of
  mirrors with historical information. Furthermore, there are
  sometimes subtle semantic differences from distribution to
  distribution, hidden behind a shared syntax. To discriminate among
  different distributions, an appropriate distribution information item
  is provided. Of course, nothing prohibits different distributions to
  agree upon the same DUDF specification instance in case they find
  that this is feasible.

\item[CUDF (Common Upgradeability Description Format)] This is the
  \emph{common} format used to abstract over distribution-specific
  details, so that solvers can be fed with upgradeability problems
  coming from any supported distribution. The CUDF format is
  specifically designed for the purpose of self-contained problem
  description.

  The conversion from a given DUDF to CUDF expands the compact
  representations that have been performed for the purpose of
  submission, exploiting distribution-specific knowledge. At the end
  of such a conversion, a problem described in CUDF is self-contained,
  only relying on the defined semantics of an upgradeability problem,
  which includes the starting state, the user query, and probably
  non-functional quality criteria.

\end{description}

\paragraph{Structure of this document} This document is structured as
follows: Chapter~\ref{chap:intro} gives introductory
information about the various kinds of documents involved in the
organization of the competition and about the problem submission
infrastructure. Chapter~\ref{chap:dudf} contains the actual
specification of the DUDF skeleton, while Chapter~\ref{chap:cudf}
contains the specification of both syntax and semantics of CUDF; both
those chapters are normative and define what it takes for a document
to be valid with respect to its specification. Appendixes to this
document contain various non-normative information, which may be
helpful to implementors of DUDF or CUDF. Documents to be made
available separately will describe how each distribution is
instantiating the DUDF skeleton.

\section{Problem data flow and submission architecture}
\label{sec:pb-dataflow}

\begin{figure}[t!]
 \begin{center}
  \includegraphics[width=0.95\textwidth]{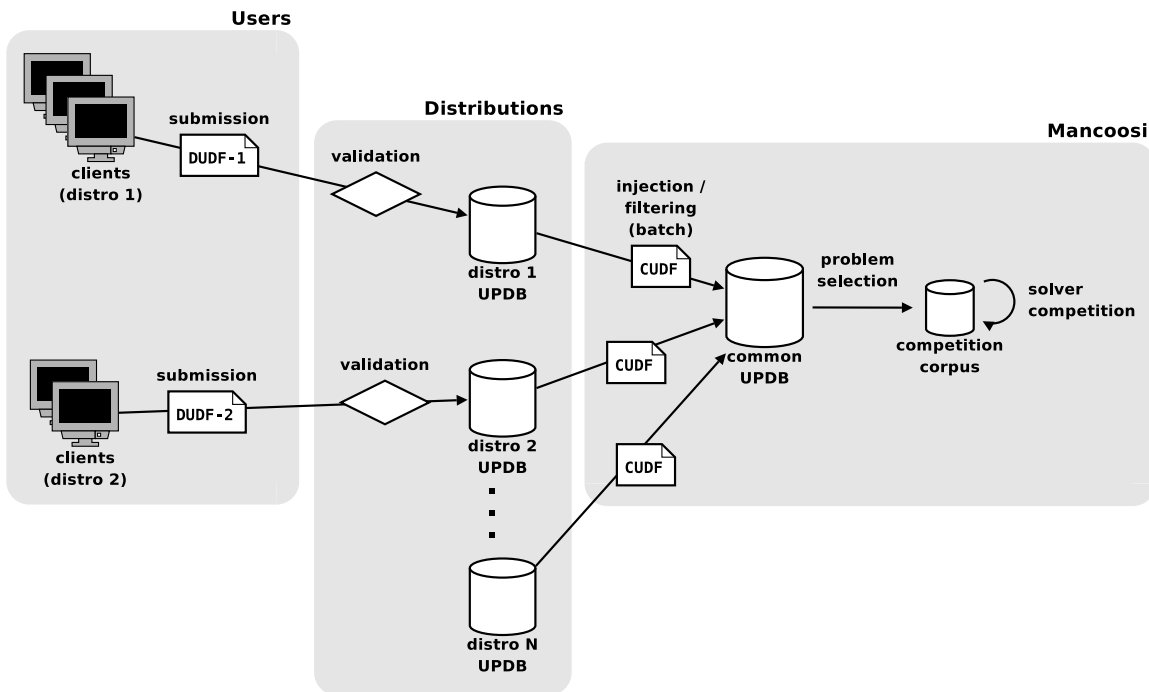}
  \caption[Problem submission data flow]{\label{fig:dataflow} Data
    flow of UPDB submissions, from users to the corpus of problems for
    the competition}
 \end{center}
\end{figure}

Figure~\ref{fig:dataflow} gives an overview of the data flow of
upgrade problems from user machines to the actual solver competition;
several stages of transmission and filtering, as well as several
different formats are involved.

Problems originate on user machines and are serialized in DUDF format
(i.e. distribution-specific DUDF instances) using some client
software. DUDF documents created that way will then be submitted to
distribution-specific repositories using some other client
software. All involved client software will be provided by
distributions, such software will constitute implementations of the
DUDF specification.

Distributions need to set up their own repositories to collect DUDF
submissions coming from their users. Submissions that do not match
the minimal quality requirements of DUDF will be rejected during a
validation phase; this mainly boils down to rejecting problems that
are not reproducible, see Chapter~\ref{chap:dudf} for more
details. All submissions that survive the validation phase are stored
by the distribution editor in a distribution-specific UPDB.

Periodically, problems collected by distributions will be injected
into a common (i.e. distribution-independent) UPDB, hosted on
an infrastructure provided by \mancoosi{} as a project resource. The
injection happens in CUDF format since distribution-specific details
are not useful for the purpose of running the competition.
Distributions are in charge of performing the conversion from DUDF to
CUDF as they are the authoritative entities for the semantics of their
proper DUDF instance and for resolving distribution-specific
references. When exactly the conversion is performed is not relevant
as long as CUDFs are ready to be injected when the periodic injections
take place.

Among all the problems collected in the common UPDB, a subset of
``interesting'' problems will then be selected to form a corpus of
problems on which the competition will be run. The act of selecting
problems will not change the document format: the resulting corpus
will still be a set of CUDF documents, chosen as a subset of the
common UPDB.

\section{Glossary}
\label{sec:glossary}

This section contains a glossary of essential terms which are used
throughout this specification.

\begin{description}
\item[Distribution] A collection of software packages that are
  designed to be installed on a common software
  platform. Distributions may come in different flavors, and the set
  of available software packages generally varies over time.
  Examples of distributions are Mandriva, Caixa Mágica, Pixart, Fedora
  or Debian, which all provide software packages for the the GNU/Linux
  platform (and probably others). The term \textit{distribution} is used
  to denote both a collection of software packages, such as
  the \textit{lenny} distribution of Debian, and the entity that
  produces and publishes such a collection, such as Mandriva,
  Caixa Mágica or Pixart. The latter are sometimes also referred to as
  \textit{distribution editors}.

  Still, the notion of distribution is not necessarily bound to FOSS
  package distributions, other platforms (e.g. Eclipse plugins, LaTeX
  packages, Perl packages, etc.) have similar distributions, similar
  problems, and can have their upgrade problems encoded in CUDF.

\item[Installer] The software tool actually responsible for
  physically installing (or de-installing) a package on a machine. This task
  particularly consists in unpacking files that come as an archive
  bundle, installing them on the user machine in persistent memory,
  probably executing configuration programs specific to that package,
  and updating the global system information on the user
  machine. Downloading packages and resolving dependencies between
  packages are in general beyond the scope of the installer. For
  instance, the installer of the Debian distribution is \texttt{dpkg},
  while the installer used in the RPM family of distributions is
  \texttt{rpm}.

\item[Meta-installer] The software tool responsible for
  organizing a user request to modify the collection of installed
  packages. This particularly involves determining the secondary
  actions that are necessary to satisfy a user request to install or
  de-install packages. To this end, a package system allows to
  declare relations between packages such as dependencies or
  conflicts. The meta-installer is also responsible for downloading
  necessary packages. Examples of meta-installers are
  \texttt{apt-get}, \texttt{aptitude} and \texttt{URPMi}.

\item[Package] A bundle of software artifacts that may be installed on
  a machine as an atomic unit, i.e. packages define the granularity
  at which software can be added to or removed from
  machines. A package typically contains an archive of files to be
  installed on a machine, programs to be executed at various stages of
  the installation or de-installation of a package, and metadata.

\item[Package status] A set of metadata maintained by the installer
  about packages currently installed on a machine. The package status
  is used by the installer as a model of the software installed on a
  machine and kept up to date upon package installation and
  removal. The kind of metadata stored for each package varies from
  distribution to distribution, but typically comprises package
  identifiers (usually name and version), human-oriented information
  such as a description of what the package contains and a formal
  declaration of the inter-package relationships of a
  package. Inter-package relationships can usually state package
  requirements (which packages are needed for a given one to
  work properly) and conflicts (which packages cannot coexist
  with a given one).

\item[Package universe] The collection of packages known to the
  meta-installer in addition to those already known to the installer,
  which are stored in the package status. Packages belonging to the
  package universe are not necessarily available on the local
  machine---while those belonging to the package status
  usually are---but are accessible in some way, for example via download from
  remote package repositories.

\item[Upgrade request] A request to alter the package status issued by
  a user (typically the system administrator) using a
  meta-installer. The expressiveness of the request language varies
  with the meta-installer, but typically enables requiring the
  installation of packages which were not previously installed, the
  removal of currently installed packages, and the upgrade to newer
  version of packages currently installed.

\item[Upgrade problem] The situation in which a user submits an
  upgrade request, or any abstract representation of such a
  situation. The representation includes all the information needed to
  recreate the situation elsewhere, at the very minimum they are:
  package status, package universe and upgrade request. Note that, in
  spite of its name, an upgrade problem is not necessarily related to
  a request to ``upgrade'' one or more packages to newer versions,
  but may also be a request to install or remove packages.  Both DUDF
  and CUDF documents are meant to encode upgrade problems for different purposes.

\end{description}

\chapter{{\em Distribution} Upgradeability Description Formats}
\label{chap:dudf}

This chapter contains the specification of the Distribution
Upgradeability Description Formats (DUDFs). Their purpose is to encode
upgrade problems as faced by users, so that they can be submitted as
candidate problems for the solver competition organized by the
\mancoosi{} project.

Additionally, DUDF can also be used as a format to store information
about the execution of a meta-installer on a user machine. A possible
use case for this is to trace information for the purpose of composing
problem reports against meta-installers. This is an added benefit for
distribution editors which is, however, beyond the scope
of the \mancoosi{} project itself.

Technically, the DUDF specification is not complete, in the sense that
some parts of DUDF documents are under-specified and called
``holes''. How to fill in those holes is a distribution-specific
decision to be taken by each distribution implementing DUDF. The
overall structure of DUDF documents is defined by the current document
and is called the \emph{DUDF skeleton}.

\section{Upgrade problems}
\label{sec:dudf-problems}

Upgrade problems manifest themselves at each attempt to change the
package status of a given machine using a meta-installer. One of the
aims of WP5 for the solver competition is to collect \emph{upgrade
  problem descriptions} which faithfully describe the upgrade problems
faced by users when invoking a meta-installer on their
machine. Informally, ``faithfully'' means that the descriptions should
contain all information needed to reproduce the problem reported by
the user, and possibly to find better solutions if they exist.

As discussed in Chapter~\ref{chap:intro}, problem descriptions will be
encoded as DUDFs and submitted to distribution-specific repositories.
Two kinds of submissions are supported by DUDF:
\begin{enumerate} \renewcommand{\labelenumi}{(\alph{enumi})}
\item Sole problem descriptions.
\item Pairs $\langle$\emph{problem description}, \emph{problem
  outcome}$\rangle$ where the outcome is a representation of the
  actual result of the originating meta-installer which has been used
  to generate the problem.
\end{enumerate}

Pairs problem/outcome are the kind of submissions to be used for the
competition. Their validity as submissions can be checked by
attempting to reproduce them upon receipt (see below), and the outcome
of competing solvers can be compared not only among each other, but
also with respect to the originating meta-installers in order to check
whether they are doing better or worse than the contenders.

Sole problem descriptions cannot be checked for reproducibility. As
such they are not interesting for the competition since they can not
be ``trusted''. Still they can be useful for purposes other than
the competition. In particular they can be used---as well as pairs
problem/outcome---by users to submit bug reports related to
installers, meta-installers, and also incoherences in package
repositories~\cite{edos-wp2d2}. This intended use is the main reason
for supporting them in this specification.

\section{Content}
\label{sec:dudf-infoitems}

A DUDF document consists of a set of \emph{information items}. Each
item describes a part of the upgrade problem faced by the user. In
this section we list the information items (or \emph{sections}) that
constitute the different kinds of DUDF submissions.

The actual format and content of each information item can either be
fully described by this specification, or be specific to some of its
instances (and hence not described here). In the latter case, we
distinguish among parts which are specific to the \emph{installer} and
parts which are specific to the
\emph{meta-installer}. Installer-specific parts have content and
format determined by the installer (e.g. rpm, dpkg, etc.) in use;
similarly, parts specific to the meta-installer are determined by the
meta-installer (e.g. apt-get, URPMi, etc.) in use.

Unless otherwise stated, all information items are required parts of
DUDF documents.

The submission of a \emph{sole upgrade problem description} consists
of the following information items:
\begin{description}

\item[Package status] (i.e. \emph{installer status}) the status of
  packages currently installed on the user machine.

  This item is installer-specific, but can also contain data specific
  to the meta-installer in case the meta-installers save some
  extended information about local packages. A concrete example of
  such extended information is the manual/automatic flag on package
  installation used by \texttt{aptitude} on Debian to implement
  ``garbage collection'' of removed packages.

\item[Package universe] the set of all packages which are known to the
  meta-installer, and are hence available for installation. This item
  is specific to the meta-installer.

  The package universe is composed of one or more \emph{package
    lists}; a number of well-known formats do exist to encode package
  lists. The package universe can generally be composed of several
  package lists, each encoded in a different format. Each package list
  must be annotated with a unique identifier describing which format
  has been used to encode the package list. A separate document will
  be published to list the set of well-known package list formats, as
  well as their unique identifiers.

\item[Requested action] the modification to the local package status
  requested by the user (e.g. ``install X'', ``upgrade Y'', ``remove
  Z''). This item is specific to the meta-installer.

\item[Desiderata] user preferences to discriminate among possible
  alternative solutions (e.g. ``minimize download side'', or ``do not
  install experimental packages''). The exact list of possible user
  preferences depends on the distribution, and on the capabilities of
  the meta-installer (for instance, for Debian's \texttt{apt} these
  may be defined in the file \path{/etc/apt/preferences}).

  This information item is optional.

\item[Tool identifiers] two pairs $\langle$\emph{name},
  \emph{version}$\rangle$ uniquely identifying the installer and
  meta-installer which are in use, in the context of a given
  distribution. One pair identifies the \emph{installer} used, the
  other the \emph{meta-installer} used.

\item[Distribution identifier] a string uniquely identifying the
  distribution run by the user (e.g. \texttt{debian},
  \texttt{mandriva}, \texttt{pixart}, \ldots), among all the
  implementations of DUDF.

  As far as GNU/Linux distributions are concerned, a hint about what to use
  as a distribution identifier comes from the file
  \path{/etc/issue}. Its content should be used as distribution
  identifier where possible.

\item[Timestamp] a timestamp (containing the same information encoded
  by dates in RFC822~\cite{rfc822} format, i.e. the same as used in
  emails) to record when the upgrade problem has been generated.

\item[Problem identifier] (i.e. \emph{uid}) a string used to identify
  this problem submission \emph{univocally}, among other submissions
  sent to the same distribution.

  The intended usage of this information item is to let CUDF documents
  cross-reference the DUDF documents which were used to generate them.
    
\end{description}

\noindent In addition to what is stated above, the submission of a pair
problem/outcome also contains the following information items:

\begin{description}

\item[Outcome] either the new local package status as seen by the used
  meta-installers (in case of \emph{success}) or an error message (in
  case of \emph{failure}, i.e. the meta-installer was not able to
  fulfill the user request).
  The error message format is specific of the used meta-installer, it
  can range from a free-text error message to a structured error
  description (e.g. to point out that the requested action cannot be
  satisfied since a given package is not available in the package
  universe).

\end{description}

It is worth noting that \mancoosi{} is not interested in all kinds of
errors, and that not all errors reported to the end user mean a
failure that is interesting for the competition. \mancoosi{} is
interested only in errors stemming from the resolution of package
relations, which is the case when the meta-installer is not able to
solve the various constraints expressed in the summary information
\emph{about} the packages. \mancoosi{} Workpackage 5 is not interested
in runtime errors such as installation failures due to disks
running out of space or execution errors of maintainer scripts. These
errors, however, may still be relevant for submitting problem reports
to a distribution vendor using the DUDF format.

Note that tool identifiers are part of the problem description since
the requested action depends on the tools the user is using. Since
available actions, as well as their semantics, can change from version
to version, tool versions are also part of the problem description.

The distribution identifier is needed to avoid bloating the
number of specified DUDFs too much. We observe that similar distributions
(e.g. Debian and Ubuntu) can submit upgrade problems using the very
same submission format (say Debian-DUDF). However, even though
extensional data (see Section~\ref{sec:dudf-intension}) are
independent of which of the similar distributions were used,
intensional data are not. Indeed, there is no guarantee that package
$p$ at version $v$ is the same on Debian and Ubuntu; similarly there
is no guarantee that an intensional package universe reference
originated on Debian is resolvable using Ubuntu historical mirrors and
vice-versa. Using the distribution identifier we can reuse the same
DUDF instance for a set of similar distributions since the
distribution identifier allows us to resolve the ambiguity.

A required property for each submission of problem/outcome pairs is
\label{req:reproduce} \emph{reproducibility}: an unreproducible submission is
useless and a waste of user bandwidth. When submissions of
problem/outcome pairs are received they have to be validated for
reproducibility. This can be achieved by keeping (possibly stripped
down) copies of commonly used tools on the server side and by running them
on the received problem description to check that the outcome matches
the reported one. Given that we are not taking into account runtime
upgrade errors, an error should manifest itself on the server side if
and only if it has manifested itself on the user machine.

\begin{figure}[t!]
 \begin{outline}
 \item dudf:
  \begin{outline}
  \item version: 1.0
  \item timestamp: \emph{timestamp}
  \item uid: \emph{unique problem identifier}
  \item distribution: \emph{distribution identifier}
  \item installer:
   \begin{outline}
   \item name: \emph{installer name}
   \item version: \emph{installer version}
   \end{outline}
  \item meta-installer:
   \begin{outline}
   \item name: \emph{meta-installer name}
   \item version: \emph{meta-installer version}
   \end{outline}
  \item problem:
   \begin{outline}
   \item package-status:
    \begin{outline}
    \item installer: \HOLE{\emph{installer package status}}
    \item meta-installer: \HOLE{\emph{meta-installer package status}}
    \end{outline}
   \item package-universe:
    \begin{outline}
    \item package-list$_1$ (format: \emph{format identifier};
      filename: \emph{path}): \HOLE{\emph{package list}}
    \item \ldots
    \item package-list$_n$ (format: \emph{format identifier};
      filename: \emph{path}): \HOLE{\emph{package list}}
    \end{outline}
   \item action: \HOLE{\emph{requested meta-installer action}}
   \item desiderata: \HOLE{\emph{meta-installer desiderata}}
   \end{outline}
  \item outcome (result: \emph{one of "success", "failure"}):
   \begin{outline} \setlength{\itemsep}{0mm}
   \item error: \HOLE{\emph{error description}} \hfill (only if result
     is ``failure'')
   \item package-status: \hfill (only if result is ``success'')
     \begin{outline}
     \item installer: \HOLE{\emph{new installer package status}}
     \item meta-installer: \HOLE{\emph{new meta-installer package
         status}}
     \end{outline}
   \end{outline}
  \end{outline}
 \end{outline}
 \caption[DUDF detailed structure]{\label{fig:dudf-outline} The DUDF
   skeleton: information items and holes corresponding to
   problem/outcome submissions.}
\end{figure}

Together, the information items supported for submissions of
problem/outcome pairs denote an outline called \emph{DUDF
  skeleton}. In the skeleton, the following information items are
\emph{holes}: package status, package universe, requested action,
desiderata and outcome. Fully determined DUDF instances are made of this
specification, together with distribution-specific documents
describing how those holes are filled. A sketch of the DUDF skeleton
is reported in Figure~\ref{fig:dudf-outline}.

Installer- or meta-installer-specific holes are denoted by framed
text. Additional information (annotations or \emph{attributes}) of
information items are reported in parentheses. The names used for
information items are for presentational purposes, yet actually
normative (see Section~\ref{sec:dudf-serialization}).

Note that in the skeleton, the package universe is sketched in its
full generality: it is made of several package lists, each of which is
annotated with its package list format. It is possible, though not
granted, that to each package list corresponds a single file on the
filesystem; in that case it is possible to annotate package lists
with a \emph{filename} containing the absolute paths corresponding to
them.

\section{Extensional vs intensional sections}
\label{sec:dudf-intension}

We have to minimize space consumption (in terms of bytes) in order not
to discourage submissions by wasting the user's resources. In
general, all the information items required for submissions are locally
available on the user machine; in principle they are all to be sent as
part of a submission. However, while some of the information items are
\emph{only} available on the user machine (e.g. current local package
status and requested action) some other items can be grouped into
parts stored elsewhere (e.g. package lists forming the current
package universe) which have possibly been replicated on the user
machine in a local cache.

We distinguish two alternative ways of sending submission
information items (or sections): a section can either be sent
intentionally or extensionally. An \emph{extensional section} is a
self-contained encoding of some information available on the user
machine, for example a dump of the current local package database, or
a dump of the current package universe.

An \emph{intentional section} is a non self-contained encoding of some
information available on the user machine, consisting of a
\emph{reference} pointing to some external resource. De-referencing
the pointer, i.e. substituting the contents of the external resource
for it, leads to the corresponding extensional section. For instance,
several distributions have package repositories available on the
Internet which are regularly updated. The current package universe
for a given user machine may correspond to package indexes downloaded
from one or several such repositories. A set of checksums of such
indexes is an example of an intensional package universe
section. Provided that a historical mirror of the distribution
repositories is available somewhere, a corresponding extensional
package universe can be built by looking up and then expanding the
checksums in the historical mirror.

The use of intensional sections instead of extensional ones is the
most straightforward space optimization we recommend to implement in
collecting problem submissions. Here are some use cases for similar
optimization:

\begin{itemize}

\item Most likely intentionality has to be used for the current
  package universe, though it will require setting up historical
  mirrors (the package metadata is sufficient for that, it will not be
  necessary to mirror the packages themselves).

\item Even though the local package status appears to be a section
  that should forcibly be sent extensionally (as the information are
  not stored elsewhere), some partial intension can be designed for
  it.

  For example, assuming that the pair $\langle pkg\_name,
  pkg\_version\rangle$ is a key univocally determining a given package
  (\emph{version uniqueness
    assumption}\label{req:versionkey}\footnote{This is assumption is
    not necessarily well-founded: users can rebuild packages locally,
    obtaining different dependency information, while retaining
    $\langle pkg\_name, pkg\_version\rangle$}), one can imagine
  sending as the local package status a set of entries $\langle
  \langle pkg\_name, pkg\_version\rangle, pkg\_status\rangle$, letting
  the server expand further package metadata (e.g. dependency
  information) on reception of the submission. In those rare cases
  where the version uniqueness assumption is not verified, the check for
  reproducibility is sufficient to spot non-reproducible submissions and
  discard them.

\item The upgrade problem outcome has to be sent extensionally as to
  check for its reproducibility upon reception. Of course, the same
  optimizations as proposed in the previous point are applicable to
  outcomes in case of success.

\end{itemize}

Any section of a submission can be sent intentionally or
extensionally, independently from the other sections; different
choices can be applied to different submissions. In fact, the choices
of how to submit the various sections are driven by the need of
fulfilling the reproducibility requirement. For instance, if a given
package universe is composed like the union of several remote package
repositories, we will need to know all the involved packages, potentially coming from any
repository in order to reproduce a submission. While a suitable intention might be
available for some repositories, this may not be the case  for some others (e.g. we might be
lacking the needed historical mirror). In such a situation the proper
solution is to send some repository reference intentionally, and the whole package listing of others
extensionally.

It is up to the DUDF submission tool to know which parts of the
package universe can be sent intentionally and which cannot.

\section{Serialization}
\label{sec:dudf-serialization}

In this section we describe how to serialize any given instance of
DUDF to a stream of bytes so that it can be serialized on disk (e.g.
to create a local archive of problem descriptions to be submitted as a
single batch) or over the network (for the actual submission to a
distribution-specific problem repository).

The serialization of DUDF is achieved by describing a mapping from the
DUDF skeleton to an XML~\cite{xmlrec} tree. The actual serialization
to bytes can then be done following the usual XML serialization rules.

To obtain the XML tree of a DUDF problem/outcome submission, one only needs to start from the corresponding outline (see
Figure~\ref{fig:dudf-outline}) and do the following:
\begin{enumerate}

\item Create a root element node called \texttt{dudf}, put it in the
  (default) namespace identified by
  \texttt{http://www.mancoosi.org/2008/cudf/dudf}.

\item Add an attribute
  \texttt{dudf:version}\footnote{The namespace prefix \texttt{dudf:}
    is bound to \texttt{http://www.mancoosi.org/2008/cudf/dudf}} to the root node, the value
  of which value is the value of the subsection \texttt{version} of the
  \texttt{dudf} section in the DUDF outline.
  
\item Starting from the DUDF outline root (and excluding the already
  processed \texttt{version} section), traverse the outline tree,
  adding child elements the general identifier of which is the section name
  used in the DUDF outline and the content of which is the result of
  recursively processing its content in the DUDF outline.
  
\item For annotated outline elements (e.g. package lists composing
  the package universe, which are annotated with format identifiers),
  map annotations to XML attributes of the relevant XML elements (note
  that the attributes should be explicitly prefixed with
  \texttt{dudf:}, as in XML attributes do \emph{not} inherit the
  default namespace).
  
\end{enumerate}

The same procedure is applied to obtain the XML tree of a DUDF sole problem submission, except that the outcome section (which should be missing anyhow in the starting DUDF outline) has to be skipped.


A non-normative example of serialization from the DUDF skeleton of
Figure~\ref{fig:dudf-outline} to XML can be found in
Appendix~\ref{chap:dudf-serialization}, Figure~\ref{fig:dudf-xml}.

\chapter{\textit{Common} Upgradeability Description Format}
\label{chap:cudf}
\label{sec:cudf-design}

This chapter contains the specification of the Common Upgradeability
Description Format (CUDF). The purpose of such a format is to encode
real upgrade problems, abstracting over details specific to a user
distribution or a package manager, so that problems coming from
different environments can be compared and treated uniformly. For the
specific purposes of \mancoosi{}, upgrade problems encoded in CUDF
format will be used to form a corpus of problems to be used in a
solver competition.

The specification of CUDF is guided by the following general design
principles:

\paragraph{Be agnostic towards distribution details}
The main purpose of CUDF, as reflected by its name, is to be a
\emph{common} format to be used to encode upgrade problems coming from
heterogeneous environments. The main environments we are considering
are FOSS distributions, but other software deployment platforms face
similar upgrade needs. As a consequence, the key design principle of
CUDF is to be agnostic with respect to distribution specific details
such as the used package system, the used installer and
meta-installer, etc. The final goal is to be able to compare problems coming from different platforms in a
uniform settings upgrade,
including at the very minimum all platforms for which a DUDF implementation (see
Chapter~\ref{chap:dudf}) has been provided.

\paragraph{Stay close to the original problem}
There are several encodings that can be considered after removing all
distribution-specific details~\cite{edos2006ase}. Since CUDF aims to
be as close as possible to the original problem we choose to avoid an
encoding where the characteristic features of the original problem are
abstracted away and are no longer distinguishable. There are various
reasons backing this choice:
\begin{itemize}

\item Preserve the ability for humans to inspect problems encoded in
  CUDF.

  Ideally, users having submitted a problem (via submission of a DUDF
  document) should be able to look at their CUDF encoding and recognize
  the upgrade situation.

\item Avoid bias towards specific upgrade planning techniques and
  implementations.

  Specific encodings (e.g. using a representation in propositional
  logic, or geared to constraint programming) bear the risk of giving
  an advantage or disadvantage to certain resolution techniques.
  Since one of our goals is to provide a set of problems to stimulate
  the advancement in upgrade planning, CUDF strives to stay
  independent of specific techniques and implementation details.

\item Make life easy to legacy tools (installers and meta-installers)
  to interact with CUDF.

  Ideally, we want legacy tools to be able to take part in the solver
  competition we are organizing. That would be easy to achieve as long
  as the CUDF encoding still resembles something with which installers
  and meta-installers are familiar. Conversely, using an encoding that
  is too abstract would constitute an obstacle for the
  state-of-the-art tools.

\end{itemize}

\paragraph{Extensibility}
CUDF has no explicit support (yet) for specifying optimization
criteria to the end of choosing the ``best'' possible solution among
all possible solutions of a given upgrade problem. The reason is that
until the end of the competition, criteria will not need to be specified
as part of the submitted problems. It would be enough to have criteria
fixed externally (e.g. the competition can have several ``tracks'',
each of which evaluates solution quality according to a single
optimization criterion), \emph{as long as} all the information needed
to evaluate the quality of a solution are encoded in CUDF.

This leads to the need of having an \emph{extensible} format to encode
upgrade problems and in particular package metadata. Indeed, since we
cannot anticipate all possible interesting optimization criteria we
can neither anticipate all the metadata that shall be stored in CUDF
documents. Hence the CUDF specification establishes a type system to be
used for typing package metadata (see Section~\ref{sec:cudf-type-ref})
and a set of core set of package metadata (see
Section~\ref{sec:cudf-properties}). Additional metadata can be added
in the future by providing their schemata, in terms of the available
types.

For example, to run a competition track in which the installed size of
all packages on the system should be minimized, the track organizers
can state that, in addition to the core package metadata, each package
must be equipped with an \texttt{Installed-Size} property, the type of which
is \texttt{posint}. The track rules will then describe how to
determine the best solution, on top of the semantics of positive
integers.

\paragraph{Transactional semantics}
Problems are encoded in CUDF according to the \emph{point of view of
  the meta-installer}, rather than to the point of view of an
installer. This means that our notion of solution correctness (see
Section~\ref{sec:cudf-semantics}) considers the resulting package
status and not \emph{how} that status is obtained on the target
machine. In particular, the order of package installations and
removals or even the various phases of package deployment and
installation (downloading, unpacking, etc.) are beyond the scope of
the CUDF encoding.

In a sense, CUDF assumes that it is possible to pass from the package
status as described in a CUDF document to any (correct) status
found by a meta-installer in a \emph{transactional} way. As an example
of a practical implication of this design principle, CUDF does not
distinguish between Debian's \texttt{Depends} and
\texttt{Pre-Depends}; note that this is coherent with the semantic
encoding of~\cite{edos2006ase}, from which the CUDF semantics takes
its inspiration.

\paragraph{Use plain text format}
On a more technical side, CUDF aims to be a \emph{simple to
  parse} (read) and \emph{simple to generate} (write) format. The reason is
as simple as our interest in providing a tool to reason about future
better upgrade planners, ignoring distracting details such as parsing
or pretty printing. Plain text is the universal encoding for
information interchange formats in the Free Software community
\cite{esr03artofunix}, using a plain text format makes it easy for
contenders to adapt tools to our format.  Moreover, it is an implicit
need if we want users to be able to ``look'' at CUDF problems and
understand them, without the need of specific tools.  Similarly, this
principle also implies that standard serialization formats should be
preferred for CUDF. In fact, the CUDF specification
describes the informative content of a CUDF document and its
semantics on one hand, and how to serialize that content to disk
(using already existing standards and technologies) on the other hand.


\section{Overview of CUDF \NONNORM}
\label{sec:cudf-overview}

This section gives an overview of the syntax and semantics of CUDF. A
precise definition of the CUDF format will be given in
Section~\ref{sec:cudf-infoitems}, while a mathematical definition of
its semantics will be given in Section~\ref{sec:cudf-semantics}. The
current section is not normative, please refer to
Sections~\ref{sec:cudf-infoitems} and~\ref{sec:cudf-semantics} for
precise definitions.

A CUDF document consists of a list of package description items, and a
user request. It is recommended that the user request be listed at the
end of the CUDF document. In the concrete representation (see
Section~\ref{sec:cudf-serialization}) each item is a stanza consisting
of one or several lines of text. It is recommended that successive
stanzas be separated by empty lines even though this is not
mandatory.

Every line in a stanza starts with the word denoting the first property
of that stanza, followed by the \LEX{:} separator and then the value
of the property, the only exception to this rule is the line
``\texttt{Problem:}'' which starts the stanza describing the query,
and which does not necessarily have a meaningful value. Other
properties of the same stanza come next, following the same
serialization conventions.

A package description stanza starts with the property \texttt{Package}
the value of which is the name of the package. Package names are
strings of length at least two, starting on a lowercase ASCII letter,
and containing only lower or uppercase ASCII letters (case is
significant), numerals, dashes \LEX{-} and dots \LEX{.}. The order of
all other properties in a package description stanza is not specified.

The only other mandatory property, besides \texttt{Package}, in a
package description stanza is \texttt{Version}, the value of which is
a positive (non-null) integer value. There may be at most one package
description stanza for any given pair of package name and version.

Then there are a number of properties that are relevant for the formal
semantics but that are only optional:
\begin{itemize}
\item The \texttt{Installed} property (the values of which are of type
  \texttt{bool}, with default value \texttt{false}) indicates whether a
  package is installed or not. It is a priori allowed to have several
  versions of the same package installed. The setting of this field in
  the stanzas of a CUDF document describe the ``initial''
  configuration of a machine, i.e., the configuration in which the
  user request is evaluated.
\item The \texttt{Keep} property has as possible values
  \texttt{version}, \texttt{package}, \texttt{feature} (being
  optional, it can also be omitted, in that case its value is
  \texttt{None}, a value shared by all omitted optional properties).
  This value is only relevant in case the \texttt{Installed} property
  is \texttt{true}. Package installations may evolve by changing the
  \texttt{Installed} property associated to pairs of package name and
  package version. The \texttt{Keep} property defines constraints on
  possible evolutions of the installation:
  \begin{description}
  \item[\tt version] means that this particular version of a package
    must not be removed,
  \item[\tt package] means that at least one version of that package
    must remained installed,
  \item[\tt feature] means that all features (see below) provided by
    this version of the package must continue to be provided,
  \item[\tt None] puts no constraint on possible evolutions of the
    installation.
  \end{description}
\end{itemize}

Then there are three properties which define relations between
packages:
\begin{itemize}
\item The \texttt{Provides} property is a possibly empty list of names
  of so-called features, also called \emph{virtual packages}. In this
  list, features may be declared either by giving an exact version, or
  without mentioning a version. Features are frequently used in
  RPM-like packaging system to declare the fact that a package
  installs a particular file on disk, and also both in RPM and
  Debian-like packaging systems to declare that a package provides a
  certain abstract functionality, like for instance
  \emph{mail-transport-agent} or \emph{postscript-reader}. A list of
  several features is interpreted as that package realizing all the
  features in the list, with the version as given in the list, or of
  \emph{all possible versions} when no particular version is mentioned
  in the list.

  The default value of that property is the empty list (that is, no
  feature is provided).

\item The \texttt{Depends} property has a complex dependency on the
  existence of packages or on features for value. Simple dependencies
  are given as the name of the package or feature, and may carry in
  addition a constraint on the version number. Version constraints can
  be of any of the form \texttt{=} $v$, \texttt{!=} $v$, \texttt{>}
  $v$, \texttt{<} $v$, \texttt{<=} $v$ or \texttt{>=} $v$ where $v$ is
  a version number. Complex dependencies are obtained by combining
  dependencies with conjunctions (denoted by \LEX{,}) and disjunctions
  (denoted \LEX{|}). However, dependencies are limited to so-called
  conjunctive normal forms, that is conjunctions of disjunctions.

  The default value of this property is the formula \texttt{True}
  (that is no particular dependency constraint).

\item The \texttt{Conflicts} property has a list of packages (or
  features), possibly equipped with package-specific version
  constraints for value; version constraints are the same as for the
  \texttt{Depends} property. Such a conflict list describes a list of
  packages that \emph{must not} be installed. For instance, if package
  $p$ of version $5$ conflicts with package \texttt{q >= 7} then we
  are not allowed to install version~$5$ of~$p$ together with any of
  the versions~$7$ or greater of~$q$. However, it would be allowed to
  install version~$5$ of~$p$ together with version~$6$ of~$q$.

  There is a special treatment for so-called \emph{self-conflicts}:
  any conflicts stemming from a pair of package $p$ and version $v$
  are ignored when checking the conflicts of this pair $(p,v)$. For
  instance, when package $p$ of version $5$ indicates that it conflicts
  with package $p$ (without version constraint) this means that
  version $5$ of package $p$ cannot be installed together with any
  other version of $p$. A conflict of package $p$ in version $5$ with
  package $p$ in version $5$ is allowed as a special case but does not
  have any effect.

  Self-conflicts of this kind are often used by packaging systems in
  order to express that only one (version of a) package implementing a
  certain feature may be installed at any given time. For instance,
  both the package \emph{sendmail} and the package \emph{postfix} (of
  any version) may provide the feature \emph{mail-transport-agent} and
  also conflict with \emph{mail-transport-agent}. The effect of this
  is that it is not possible to install \emph{sendmail} and
  \emph{postfix} together (or any of them together with any other
  package providing \emph{mail-transport-agent}), but it does allow to
  install \emph{sendmail} or \emph{postfix} since the conflict
  stemming from the package itself is ignored.

  The default value of this property is the empty list (that is, no
  conflict declared here).
\end{itemize}

Finally, the CUDF document contains a stanza representing the user
request.  This stanza starts with the line \texttt{Problem:}, and it
may contain an \texttt{Install} property, a \texttt{Remove} property,
and a \texttt{Upgrade} property. Each of these properties is optional,
their value is a list of packages (or features) possibly equipped with
version constraints; the default value of these three properties is
the empty list. The \texttt{Install} property gives packages that are
demanded to be installed, while the \texttt{Remove} gives packages
that must be removed. The \texttt{Upgrade} property has a similar
meaning as \texttt{Install}, the difference being that the former
requires that for every package in that list only one version be
installed, and that this version be greater or equal to any version of
that package previously installed.

\section{Content}
\label{sec:cudf-infoitems}

A CUDF document (or simply ``CUDF'') is composed of a set of
\emph{information items}. Each item represents a part of the original
upgrade problem.

Each information item belongs exactly to one of the following two classes:
\begin{description}

\item[Package description items] specify packages that have a role in
  the upgrade problem described by a given CUDF.

  A CUDF document contains several package description information
  items. In a typical scenario there is one such item for each package
  known to the package manager, including both locally installed
  packages (as part of the local status) and packages available from
  remote repositories known to the meta-installer (as part of the
  package universe).

\item[Problem description items] describe other information items that
  contribute to create the upgrade problem, but which are not specific
  to any particular package. A CUDF document must contain exactly one
  problem description item.

  Intuitively, the item contains global information about the upgrade
  problem. At the very minimum, that information contains the request
  submitted by the user to the package manager.

\end{description}

\begin{figure}[t!]
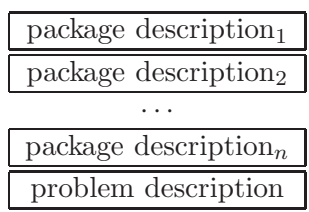

  \begin{center}
    \begin{tabular}{|c|}
      \hline
      package description${}_1$ \\
      \hline\hline
      package description${}_2$ \\
      \hline
      \multicolumn{1}{c}{$\cdots$} \\
      \hline
      package description${}_n$ \\
      \hline\hline
      problem description \\
      \hline
    \end{tabular}
  \end{center}
  \caption[CUDF overall structure]{\label{fig:cudf-items} Overall
    structure of a CUDF document; information items are represented
    according to the canonical CUDF ordering.}
\end{figure}

CUDF documents consist of a single problem description item and
several package description items.\footnote{There is no constraint on
  the number of package description items, but problems represented as
  CUDFs with no package description items are uninteresting. CUDFs are
  expected to include at least one package description item, and
  usually many more than just one.} While there is no strict
imposition on the relative order of information items in actual CUDFs,
this specification assumes the \emph{canonical ordering} of having first all package descriptions and
then the single problem description, for uniformity of presentation. A schematic representation of a
typical CUDF and its canonical ordering is given in
Figure~\ref{fig:cudf-items}. A similar, yet more detailed, pictorial
representation of CUDFs is given later on in
Figure~\ref{fig:cudf-details}.

CUDF implementations should implement the canonical ordering whenever
possible.

\subsection{Generalities}

Each information item consists of a set of \emph{properties}. Each
property has a name and a value (i.e. each property consists of a
name-value pair).

A \emph{property name} is a string of Unicode~\cite{unicode}
characters matching the additional constraint of being an
identifier. An \emph{identifier} is a non-empty string composed only
of characters belonging to the following character classes:
\begin{description}
\item[Lowercase Latin letters] from ``\verb+a+'' (Unicode code point
  \texttt{U+0061}) to ``\verb+z+'' (\texttt{U+007A}), in the ordering
  induced by Unicode code points.
\item[Uppercase Latin letters] from ``\verb+A+'' (\texttt{U+0041}) to
  ``\verb+Z+'' (\texttt{U+005A}).
\item[Arabic numeral digits] from ``\verb+0+'' (\texttt{U+0030}) to
  ``\verb+9+'' (\texttt{U+0039}).
\item[Separators] the character ``\verb+-+'' (\texttt{U+002D})
\end{description}
Additionally, identifiers must start with one of the lowercase or
uppercase Latin letters defined above.

A \emph{property value} is a typed value, belonging to some set. We
call this set the \emph{set of values} or the \emph{type domain} of the
type associated to a property. The type is fixed for each property:
any given property can only assume values having the very same type
and coming from the very same set of values; the description of each
supported property in this specification states what is the type of
its values.

A property can be either optional or required. A property is optional
if its value is indicated as optional in the property description,
otherwise it is required. Required properties must always be present as
part of the information items they belongs to, while optional
properties may not be present.  Optional properties that are present
in CUDFs must assume a value belonging to its type domain. Optional
properties can have a default value; it must be a value belonging to
its type domain.

Optional properties that are not present in CUDFs and have a default
value $v$ are treated as properties assuming value $v$. It is
indistinguishable whether the value was actually specified in the CUDF
serialization or not. Optional properties that are not present in
CUDFs and do not have an optional value are treated as properties
assuming the distinguished value \texttt{None}. The value
\texttt{None} does not belong to any set of values which can be
written in CUDF serializations; this feature allows us to distinguish
whether an optional property has been specified in actual
CUDFs or not.

Each property supported by CUDF can be fully specified using a
\emph{property schema}. Such a schema consists of:
\begin{itemize}
\item the name of the property;
\item the type of property values;
\item the information item the property belongs to;
\item the optionality of the property (i.e. whether the described
  property is required in each instances of the information item it
  belongs to), optionality is either ``required'' or ``optional'';
\item for optional properties only, an optional default value.  It is
  possible to give a default value for an optional property, but is
  not mandatory to do so.
\end{itemize}

Actual CUDF documents must contain all
required properties for each information item. For both required and optional properties, the
type of property values must match the type prescribed by property
schemata.

Section~\ref{sec:cudf-properties} gives the schemata of the \emph{core
property set} supported by CUDF. Nevertheless the set of properties
which can be given to build information items is open-ended
(open-world assumption), and not restricted to the core set.
Information items can contain extra properties \emph{not} prescribed by
this specification as long as their names do not clash with names of
properties in the core set. It is up to implementations to make use of
such extra properties, to define their names and the type of their
values. Of course all extra properties are optional as far as
conformance to this specification is concerned.

\subsection{Types}
\label{sec:cudf-types}

As discussed above, each property value has a type which is fixed for
any given property. A \emph{type} is a set of values, which is also
called \emph{value space} or \emph{domain} of a given type. Let $t$ be
a type, we denote with $\mathcal{V}(t)$ its value space. Moreover,a \emph{lexical space} $\mathcal{L}(t)$
 is associated to each type, and it denotes the set of possible representations of all values belonging
to the value space as strings of Unicode characters. The relationships
between the value spaces and lexical spaces are as follows:
\begin{itemize}
\item For each $l \in \mathcal{L}(t)$ there is a unique corresponding
  value $\mathit{parse}_t(l) \in \mathcal{V}(t)$. The function
  $\mathit{parse}_t(\cdot)$ is the \emph{parsing} (partial) function
  used to parse syntactic values into semantic values.

\item For each $v \in \mathcal{V}(t)$ there can be several $l \in
  \mathcal{L}(t)$ such that $v = \mathit{parse}_t(l)$, i.e. the
  parsing function is not necessarily one-to-one.
\end{itemize}


\paragraph{Subtyping}
Interesting relationships also exist between types, in particular
\emph{subtyping}. A type $t_2$ is said to be a subtype of a
\emph{supertype} $t_1$ (written $t_2\SUBTY t_1$) if $\mathcal{V}(t_2)
\subseteq \mathcal{V}(t_1)$, that is, when the domain of the subtype
is contained in the domain of its supertype.  Given $t_2\SUBTY t_1$,
the lexical space of $t_2$ can be obtained by restricting the lexical
space of $t_1$ to all elements which can be parsed to elements of the
value space of $t_2$, i.e. $\mathcal{L}(t_2) = \{ l\in
\mathcal{L}(t_1) ~|~ \mathit{parse}_{t_1}(l) \in \mathcal{V}(t_2)
\}$. Therefore the parsing function for a given subtype can be
obtained by simply reusing the parsing function of the supertype
treating as parsing errors all values not belonging to the domain of
the subtype.

As a consequence of the above definitions and properties, each type
can be defined by describing its value and lexical spaces, as well as
the semantics of its parsing functions. Subtypes can be defined by
simply giving restrictions on the value space of supertypes. The
section further gives the definitions for all types used by
CUDF.

\paragraph{Conventions}
In this specification abstract values belonging to the value space are
denoted using mathematical notation.

Lexical values are denoted by double-quoted strings typeset in
\texttt{monospace font} and encoded in UTF-8. The double-quotes are
used for presentational purposes of this specification and are not
part of the actual lexical value. Such a value can be found by
considering the Unicode string corresponding to the given UTF-8
string, after having removed double quotes. For example, the lexical
value \LEX{foo} denotes the Unicode string of length 3, composed of
the three lowercase letters ``f'' (Unicode code point \texttt{U+0066}),
``o'' (\texttt{U+006F}), and ``o'' again.

Functions can either be described intentionally or extensionally. In
the former case, types are specified via natural language explanation
of their semantics, or reference to functions described elsewhere. In
the latter case they are defined by enumerating argument/result pairs
using the following notation: $\{\mathit{input}_1\to\mathit{output}_1,
\ldots, \mathit{input}_n\to\mathit{output}_n\}$.

For the sake of brevity, several details about lexical values
and parsing functions are deferred to external specifications, most
notably to ``XML Schema Part 2: Datatypes''~\cite{xmlschema-part2},
which specify a set of simple datatypes, providing for each of them
notions similar to the one introduced above: value space, lexical
space and parsing functions. When deferring a definition to the
definition of the corresponding XML Schema datatype, we will write
\XSTYPE{xs:foo}, where ``xs:foo'' is the XML Schema datatype name.

Complex lexical spaces are sometimes described by the means of EBNF
grammars~\cite{ebnf} which use \textsc{SmallCaps} for non terminal
symbols and double-quoted string as described above for terminals.
Grammars are always given with the productions of their start symbol
first. In order to avoid duplications, grammars appearing later on in
this specification can reuse symbols defined in previous grammars.
When EBNF grammars are used, the definition of parsing functions can
be omitted and delegated to parsers built using the given grammar. For
the completeness of this specification it is enough to state how the
values associated to non terminals have to be translated to elements
in the value space (i.e. to give the ``semantic actions'' associated
to grammar productions).

\subsubsection{CUDF type library}
\label{sec:cudf-type-ref}

In the presentation of the available CUDF types that follows, we first
introduce all \emph{primitive types}, i.e. all those types that are
not obtained via subtyping; then we describe \emph{derived types},
i.e. those that are obtained as subtypes of other (primitive or
derived) types. As discussed above, each derived type can be described
by simply giving a restriction of the value space of its supertype.

\begin{figure}[t!]
  \begin{center}
    \includegraphics[width=0.90\textwidth]{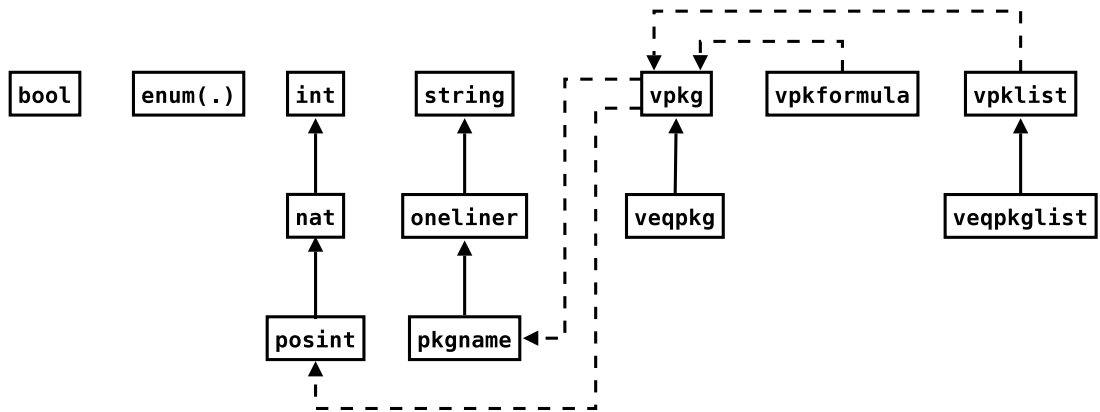}
  \end{center}
  \caption[CUDF types and their relationships.]{\label{fig:cudf-types}
    CUDF types and their relationships.}
\end{figure}

Figure~\ref{fig:cudf-types} shows a diagram giving an overview of CUDF types and their relationships. In the figure, directed straight
arrows denote subtyping relationships, with the type pointed at being
the supertype; directed dashed arrows denote acquaintenance, i.e. the
fact that the values of one type contain values of other types, the
latter being pointed at by the arrow. Transitive relationships are
omitted from the figure.

\TYPEBOX{bool}
        {Boolean values}
        {The set of distinguished values $\{\mathtt{true}, \mathtt{false} \}$}
        {The set of strings $\{$ \LEX{true}, \LEX{false}, $\}$}
        {$\{\LEX{true}\to\mathtt{true}, \LEX{false}\to\mathtt{false}\}$}

\TYPEBOX{int}
        {Integer numbers}
        {The set of integer numbers $\mathbb{Z} = \{\ldots, -2, -1, 0,
          1, 2, \ldots\}$ (Note that this set is infinite, unlike
          legacy integers available in most programming languages.)}
        {The same lexical representation as the one used by the
          \XSTYPE{xs:integer}, i.e. finite-length sequences of
          decimal digits (\texttt{U+0030}--\texttt{U+0039}) with an
          optional leading sign (defaulting to \LEX{+}). For example:
          \LEX{-1}, \LEX{0,} \LEX{12678967543233}, \LEX{+100000}.}
        {The same parsing function as the one used for \XSTYPE{xs:integer}}

\TYPEBOX{string}
        {Unicode strings}
        {The set of---possibly empty---strings of arbitrary Unicode
          characters.}
        {Some (specified) character encoding supported by Unicode. For
          the purpose of CUDF serialization the character encoding of
          choice is UTF-8 (see Section~\ref{sec:cudf-serialization}).}
        {The same parsing functions as the ones used for \XSTYPE{xs:string}, i.e.
          simply decoding from the used character encoding to Unicode
          character strings.}

We assume the notion of the function \emph{length} for Unicode strings,
which is defined by counting the number of Unicode characters (not
bytes) forming a Unicode string in a given encoding.  As a
consequence, the empty string \LEX{} has length 0.

\TYPEBOX{vpkg}
        {Versioned package names}
        {The set of pairs $\langle \mathit{vpred}, p\rangle$ where $p$
          is a value of type \texttt{pkgname} (see below) and
          $\mathit{vpred}$ is either $\top$ (denoting that no version
          constraint has been imposed on package name $p$) or a pair
          $\langle \mathit{relop}, v\rangle$ (denoting that a specific
          version constraint is in effect on package name $p$). In the
          latter case $\mathit{relop}$ is one of $\{\mathtt{=},
          \mathtt{\neq}, \mathtt{\geq}, \mathtt{>}, \mathtt{\leq},
          \mathtt{<}\}$ and $v$ is a value of type \texttt{posint}
          (see below).}
        {The set of strings matching the grammar:
          \[
          \begin{array}{rcl}
            \NT{VPkg} & ::= & \NT{PkgName} ~\NT{Sp}+ ~\NT{VConstr}? \\
            \NT{VConstr} & ::= & ~\NT{Sp}* ~\NT{RelOp}
             ~\NT{Sp}+ ~\NT{Ver} ~\NT{Sp}* \\
            \NT{RelOp} & ::= & \LEX{=} ~|~ \LEX{!=} ~|~ \LEX{>=} ~|~
             \LEX{>} ~|~ \LEX{<=} ~|~ \LEX{<} \\
            \NT{Sp} & ::=
             & \LEX{ } \mbox{\hspace{1cm}(i.e. \texttt{U+0020})} \\
            \NT{Ver} & ::= & \NT{PosInt} \\
          \end{array}
          \]
          where the nonterminal \NT{PkgName} matches lexical values
          of \texttt{pkgname} (see below) and \NT{PosInt} those of
          \texttt{posint} (see below). The values resulting from
          parsing \NT{VConstr}, which match \NT{RelOp} and
          \NT{Version} respectively, are used to form the internal
          pair $\langle \mathit{relop}, \mathit{v}\rangle$; similarly,
          the values resulting from parsing \NT{VPkg} are used to form
          the external pair $\langle \mathit{vpred},
          p\rangle$.}
        {Induced by the grammar.

          \NT{RelOp} is parsed by the function: $\{
          \LEX{=}\to\mathtt{=}, \LEX{!=}\to\mathtt{\neq},
          \LEX{>=}\to\mathtt{\geq}, \LEX{>}\to\mathtt{>},
          \LEX{<=}\to\mathtt{\leq}, \LEX{<}\to\mathtt{<}\}$.}

The semantics of versioned package names depend on the context where
they appear. Generally, package names without version constraints are
to be intended as package predicates matching all packages with a
given name. Package names with a version constraint will
additionally satisfy the given version requirement.

\TYPEBOX{vpkgformula}
        {Formulae over versioned package names}
        {The smallest set $F$ such that:
          \[
          \begin{array}{lll}
            \mathtt{true} \in F & & \mbox{(truth)} \\
	    \mathcal{V}(\mathtt{vpkg}) \subseteq F & & \mbox{(atoms)} \\
            \bigvee_{i=1,\ldots,n} a_i \in F
             & a_1,\ldots,a_n \mbox{ atoms} \in F
             & \mbox{(disjunctions)} \\
            \bigwedge_{i=i,\ldots,n} d_i \in F
             & d_1,\ldots,d_n \mbox{ disjunctions} \in F
             & \mbox{(conjunctions)} \\
          \end{array}
          \]}
        {The set of strings matching the following grammar:
          \[
          \begin{array}{rcl}
            \NT{Fla} & ::= & \NT{AndFla} \\
            \NT{AndFla} & ::= & \NT{OrFla} ~(\NT{Sp}* ~\LEX{,}
             ~\NT{Sp}* ~\NT{OrFla})* \\
            \NT{OrFla} & ::= & \NT{AtomFla} ~(\NT{Sp}* ~\LEX{|}
             ~\NT{Sp}* ~\NT{AtomFla}) \\
            \NT{AtomFla} & ::= & \NT{VPkg} \\
          \end{array}
          \]}
        {Induced by the grammar.

          \NT{AtomFla} nonterminals are parsed as atoms, \NT{OrFla}
          as disjunctions of the atoms returned by their
          \NT{AtomFla}s, \NT{AndFla} as conjunctions of the
          disjunctions returned by their \NT{OrFla}s.}

Note that formulae over versioned package names are always in
conjunctive normal form (CNF), i.e. they always have the shape of
``conjunctions of disjunctions of atomic formulae''.

\TYPEBOX{vpkglist}
        {Lists of versioned package names}
        {The smallest set $L$ such that:
          \[
          \begin{array}{lll}
            \mathtt{[]} \in L & & \mbox{(empty lists)} \\
            \mathit{p} \mathtt{::} l \in L
             & \mathit{p} \in \mathcal{V}(\mathtt{vpkg}), l \in L
             & \mbox{(package concatenations)} \\
          \end{array}
          \]}
        {The set of strings matching the grammar:
          \[
          \begin{array}{rcl}
            \NT{VPkgs} & ::= & \LEX{} ~|~ \NT{VPkg} ~(\NT{Sp}* ~\LEX{,}
             ~\NT{Sp}* ~\NT{VPkg})*\\
          \end{array}
          \]}
        {Induced by the grammar.

          \LEX{} is parsed as $\mathtt{[]}$, while an instance of
          \NT{VPkg} followed by a list of versioned package names is
          parsed as package concatenation.}

\TYPEBOX{enum($s_1,\ldots,s_n$)}
        {Enumerations}
        {Rather than a single type, \texttt{enum} is a type scheme
          defining infinite possible actual types. All those types are
          indexed by the set of symbols $\{s_1,\ldots,s_n\}$, for any
          such set a single type (an \emph{enumeration}) is defined.

          Each enumeration is a type, its values can be one of the
          symbols $s_1, \ldots, s_n$. Symbols must match the
          constraints of identifiers. For convenience, in this
          specification symbols are written as strings, but without
          the external double quotes, and prefixed by a single quote
          \texttt{'}.}
        {$\{s\in\mathcal{L}(\mathtt{string}) ~|~ s \mbox{ is an
            identifier}\}$}
        {$\{\mbox{\tt"}s\mbox{\tt"}\to\mbox{\tt'}s ~|~ s \mbox{ is an
            identifier}\}$

          The parsing function is defined point-wise on each Unicode
          string matching the constraints of identifiers. For each of
          them, the parsing function returns a symbol, the name of which is
          that very same identifier.}

For example, given the enumeration $E=\mathtt{enum}(\mathtt{'foo},
\mathtt{'bar}, \mathtt{'baz})$, we have the following: $\mathtt{'foo}\in
\mathcal{V}(E)$, $\mathtt{'bar}\in \mathcal{V}(E)$, and
$\mathtt{'baz}\in \mathcal{V}(E)$. Note that \texttt{None} is not part
of any enumeration, but optional properties having enumeration types
can assume the \texttt{None} value as usual.

\TYPEBOX{oneliner}
        {One-liner Unicode strings}
        {\texttt{oneliner} is a subtype of \texttt{string}.

          The set of, possibly empty, strings of Unicode characters
          not containing any of the following ``newline'' characters:
          line feed (\texttt{U+000A}), carriage return
          (\texttt{U+000D}).

          Note that, in spite of \XSTYPE{xs:string} not being a type
          available for use in CUDF, CUDF \texttt{oneliner} is
          conceptually a subtype also of \XSTYPE{xs:string} obtained
          by removing newline characters from its value space.}
        {As per subtyping.}
        {As per subtyping.}

\TYPEBOX{pkgname}
        {Package names}
        {\texttt{pkgname} is a subtype of \texttt{oneliner}.

          It is obtained allowing only strings
          that satisfy the following condition in the value space:
          \begin{itemize}
          \item the string starts with a lowercase Latin letter
          \item the string only consists of: lowercase Latin
            letters, Arabic numeral digits, dashes (\texttt{U+002D}),
            dots (\texttt{U+002E})
          \item the string has length 2 or greater
          \end{itemize}}
        {As per subtyping.}
        {As per subtyping.}

\TYPEBOX{nat}
        {Natural numbers}
        {\texttt{nat} is a subtype of \texttt{int}.

          It is obtained by allowing only non-negative integers in the
          value space.}
        {As per subtyping.}
        {As per subtyping.}

\TYPEBOX{posint}
        {Positive natural numbers}
        {\texttt{posint} is a subtype of \texttt{nat}.

          It is obtained by removing the number $0$ from the value
          space of \texttt{nat}.}
        {As per subtyping.}
        {As per subtyping.}

\TYPEBOX{veqpkg}
        {Version-specific package names}
        {\texttt{veqpkg} is a subtype of \texttt{vpkg}.

          It is obtained by removing all packages with version constraints other
          than $=$, more formally: $\mathcal{V}(\mathtt{veqpkg}) =
          \{\langle\mathit{vpred}, p\rangle ~|~ \langle\mathit{vpred},
          p\rangle \in \mathcal{V}(\mathtt{vpkg}), \mathit{vpred} =
          \top \lor \mathit{vpred} = \langle \mathtt{=}, v \rangle
          \mbox{ for some } v\}$ from the value space of
          \texttt{vpkg}.}
        {As per subtyping.}
        {As per subtyping.}


\TYPEBOX{veqpkglist}
        {Lists of version-specific package names}
        {\texttt{veqpkglist} is a subtype of \texttt{vpkglist}.

          It is obtained by using as value space only the smallest set
          $L'\subseteq \mathcal{V}(\mathtt{vpkglist})$ such that:
          \[
          \begin{array}{lll}
            \mathtt{[]} \in L' & & \mbox{(empty lists)} \\
            \mathit{p} \mathtt{::} l \in L'
             & \mathit{p} \in \mathcal{V}(\mathtt{veqpkg}), l \in L'
             & \mbox{(package concatenations)} \\
          \end{array}
          \]}
        {As per subtyping.}
        {As per subtyping.}

\subsection{Property schemata}
\label{sec:cudf-properties}

Each of the information items supported by CUDF (either package or
problem description items, see Section~\ref{sec:cudf-infoitems}) is
composed of several properties. In this section we give the schemata
for all properties that can be part of package description items and
problem description items.

\subsubsection{Package description}
\label{sec:cudf-packages}

A package description item describes several facets of a package.

\PROPERTYBOX{Package}
            {pkgname}
            {required}{}
            {Name of the package being described.}

\PROPERTYBOX{Version}
            {posint}
            {required}{}
            {Version of the package being described.}

\PROPERTYBOX{Depends}
            {vpkgformula}
            {optional}{$\mathtt{true}$}
            {Intentional representation of the dependencies of the
              package being described.}

Dependencies indicate which packages need to be installed to make a
given package installable. Dependencies are indicated as
boolean CNF formulae over possibly versioned package
names. Dependencies are the most expressive relationships which can be
stated among packages using CUDF properties.

\PROPERTYBOX{Conflicts}
            {vpkglist}
            {optional}{$\mathtt{[]}$}
            {Intentional representation of packages which conflict
              with the package being described.}

Conflicts indicate which packages cannot be co-installed, in any given
installation, together with a given package. Note that the language to
express conflicts is more limited than that used to express
dependencies: it consists of plain lists of possibly versioned package
names, rather than CNF formulae.

Also note that as far as CUDF is concerned there are no implicit conflicts
assumed between different versions of the same package, if they are
intended they need to be explicitly specified using the
\texttt{Conflicts} property. According to the CUDF semantics this can
be achieved by declaring, for a package $p$, a conflict with $p$
itself; see Section~\ref{sec:cudf-semantics} for more information.

\PROPERTYBOX{Provides}
            {veqpkglist}
            {optional}{$\mathtt{[]}$}
            {Features provided by the package being described.}

A package can declare zero or more \emph{features} that it provides.
To abstract over package names, other packages may declare
relationships with such features. Packages can provide a specific
version of a given feature, or provide a feature without mentioning a
version (meaning that \emph{all} possible versions of a given feature
are provided by that package).

\PROPERTYBOX{Installed}
            {bool}
            {optional}{$\mathtt{false}$}
            {Flag stating whether or not the package being described
              is installed.}

Two kinds of packages play a role in the upgrade process: currently
installed packages constituting the local package status and (locally
or remotely) available packages which are known to the meta-installer
and constitute the package universe. \texttt{Installed} distinguishes
among these two cases, it is \texttt{true} for packages which are part
of the local status and \texttt{false} for those which are part of the
package universe. Other kinds of packages that do not play a role in
the package upgrade problem are not represented in CUDF.

\PROPERTYBOX{Keep}
            {enum$(\mathtt{'version}, \mathtt{'package},
              \mathtt{'feature})$}
            {optional}{}
            {Indication of which aspects of the package being
              described the user wants to preserve across upgrades.
              \begin{itemize}
              \item \texttt{'version} means preserving the current
                version, as recorded in the package status.
              \item \texttt{'package} means preserving at least one
                version of the package in the resulting package
                status.
              \item \texttt{'feature} means preserving all the
                provided features.
            \end{itemize}}

Note that it is not specified \emph{how} the requirements of the
\texttt{Keep} feature have to be fulfilled; in the particular case of
\texttt{'feature} it is possible that the requirement gets fulfilled
by replacing a package by some other packages, which, together,
provide the same features. See Section~\ref{sec:cudf-semantics} for
the formal specification of the meaning of the \texttt{Keep} property.

\subsubsection{Problem description}
\label{sec:cudf-global}

\PROPERTYBOX{Install}
            {vpkglist}
            {optional}{$\mathtt{[]}$}
            {List of packages the user wants to be installed.}

\PROPERTYBOX{Remove}
            {vpkglist}
            {optional}{$\mathtt{[]}$}
            {List of packages the user wants to be removed.}

\PROPERTYBOX{Upgrade}
            {vpkglist}
            {optional}{$\mathtt{[]}$}
            {List of packages the user wants to be upgraded to newer
              versions.}

The properties \texttt{Install}, \texttt{Remove} and \texttt{Upgrade}
provide the same mechanism for specifying the target packages: lists
of package names with optional version specifications. A properly
completed \texttt{Install} action ensures that the requested packages
are installed in the resulting package status, on the contrary
\texttt{Remove} ensures that they are not. Since CUDF supports multiple installed versions of the same package in
principle there is no
implicit need of removing other packages due to homonym upon
\texttt{Install}. \texttt{Upgrade} is similar to \texttt{Install}, but
additionally ensures that only one version of each of the target
packages is preserved in the resulting packages status; it also
ensures that newer versions of them get installed. See
Section~\ref{sec:cudf-semantics} for a formal specification of the
semantics of actions.

\subsection{Document structure}
\label{sec:cudf-alltogether}

\begin{figure}[t!]
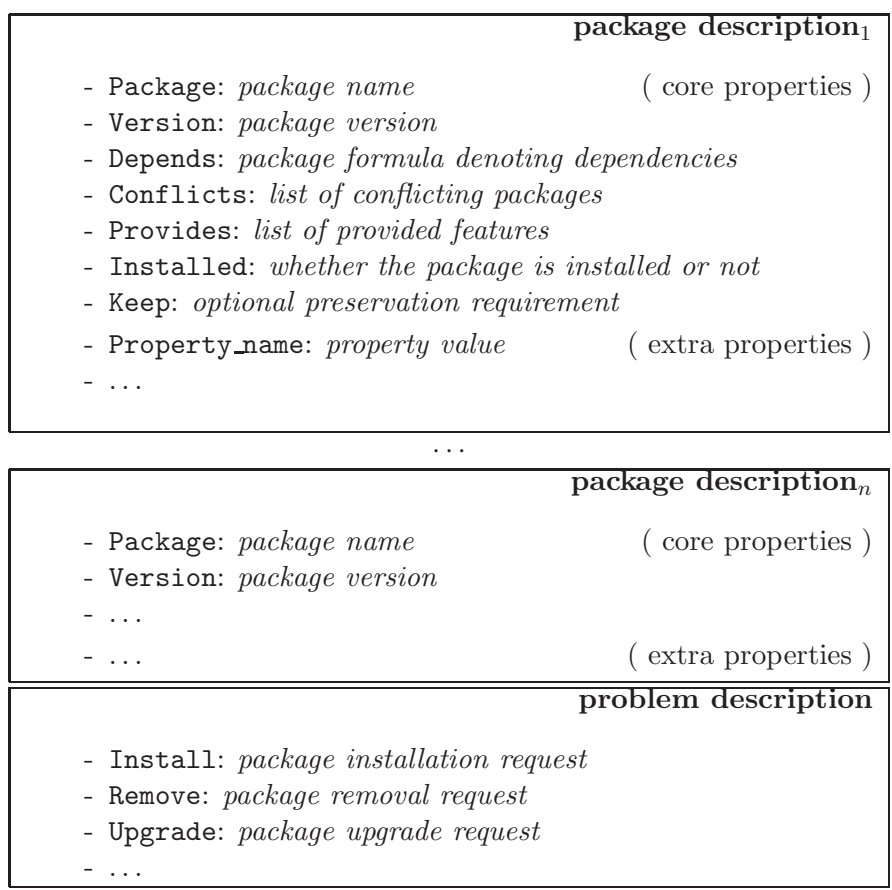

  \begin{center}
    \begin{tabular}{|c|}
      \hline

      \begin{minipage}{0.7\textwidth}
        ~ \hfill \textbf{package description${}_1$}
        \begin{outline}
        \item \texttt{Package}: \emph{package name} \hfill ( core
          properties )
        \item \texttt{Version}: \emph{package version}
        \item \texttt{Depends}: \emph{package formula denoting
          dependencies}
        \item \texttt{Conflicts}: \emph{list of conflicting packages}
        \item \texttt{Provides}: \emph{list of provided features}
        \item \texttt{Installed}: \emph{whether the package is
          installed or not}
        \item \texttt{Keep}: \emph{optional preservation requirement}
        \end{outline}
        \vspace{-4ex}
        \begin{outline}
        \item \texttt{Property\_name}: \emph{property value} \hfill (
          extra properties )
        \item \ldots{}
        \end{outline}
        ~
        \vspace{-2ex}
      \end{minipage} \\

      \hline
      \multicolumn{1}{c}{$\cdots$} \\
      \hline

      \begin{minipage}{0.7\textwidth}
        ~ \hfill \textbf{package description${}_n$}
        \begin{outline}
        \item \texttt{Package}: \emph{package name} \hfill ( core
          properties )
        \item \texttt{Version}: \emph{package version}
        \item \ldots
        \end{outline}
        \vspace{-4ex}
        \begin{outline}
        \item \ldots \hfill ( extra properties )
        \end{outline}
        ~
        \vspace{-4ex}
      \end{minipage} \\

      \hline\hline
      \begin{minipage}{0.7\textwidth}
        ~ \hfill \textbf{problem description}
        \begin{outline}
        \item \texttt{Install}: \emph{package installation request}
        \item \texttt{Remove}: \emph{package removal request}
        \item \texttt{Upgrade}: \emph{package upgrade request}
        \item \ldots
        \end{outline}
        ~
        \vspace{-4ex}
      \end{minipage} \\
      \hline
    \end{tabular}
  \end{center}
  \caption[CUDF detailed structure]{\label{fig:cudf-details} Detailed
    structure of a CUDF document with highlight of core properties.}
\end{figure}

Putting it all together, the detailed structure of CUDF document is as
depicted in Figure~\ref{fig:cudf-details}; the figure has to be
interpreted as a refined version of Figure~\ref{fig:cudf-items}, which
we are now able to fill with the properties described in the previous
section. Note that all core properties are shown in the figure, in
spite of their optionality.

\subsubsection{Global constraints}

In addition to the per-property constraints reported in the previous
section, CUDF documents must respect extra constraints which are not
specific to sole information items or properties.

\begin{description}

\item[Package/version uniqueness] among all package description items
  forming a given CUDF, there must not exist two package descriptions
  $p_1$ and $p_2$ such that they have the same value of the property
  \texttt{name} and the same value of the property \texttt{version},
  i.e. the pair of property values $\langle \mathtt{name},
  \mathtt{version}\rangle$ can be used as a ``key'' to look up package
  descriptions in a given CUDF.


\end{description}

There is no strict imposition neither in specifying at least one of
the \texttt{Install}/\texttt{Remove}/\texttt{Upgrade} properties, nor
in specifying non empty-lists as their values. Nevertheless, CUDFs
representing problems with empty queries are mostly uninteresting.

\section{Formal semantics}
\label{sec:cudf-semantics}

The semantics is defined in a style similar to~\cite{edos2006ase},
however, we now have to deal with an abstract semantics that is closer
to ``real'' problem descriptions, and that contains artifacts like
\emph{features}. This induces some complications for the definition of
the semantics. In \cite{edos2006ase} this and similar problems were
avoided by a pre-processing step that expands many of the notions that
we wish to keep in the CUDF format.

\subsection{Abstract syntax and semantic domains}
The abstract syntax and the semantics is defined using the value domains
defined in Section~\ref{sec:cudf-types}. In addition, we give the following
definitions:

\begin{definition}
  \label{def:domains:basic}
  \begin{itemize}
  \item \DomConstraints{} is the set of version constraints,
    consisting of the value $\top$ and all pairs $(\textit{relop},v)$
    where \textit{relop} is one of $=,\neq,<,>,\leq,\geq$ and
    $v\in\DomVersions$.

  \item \DomKeep{} is the set of the possible values of the
    \texttt{Keep} property of package information items, that is: $\{
    \texttt{'version}, \texttt{'package}, \texttt{'feature},
    \texttt{None} \}$
  \end{itemize}
\end{definition}

The abstract syntax of a CUDF document is a pair consisting of a
package description (as defined in
Definition~\ref{def:package:description}) and a request (see
Definition~\ref{def:request}).

\begin{definition}[Package description]
  \label{def:package:description}
  A \emph{package description} is a partial function
  \begin{eqnarray*}
  \DomPackages\times\DomVersions & \rightsquigarrow \\
  &&\hskip -4cm
  \DomBool \times \DomKeep\times \DomForm \times \DomCList \times \DomEList
  \end{eqnarray*}
  The set of all package descriptions is noted \PackageDescriptions.
  If $\phi$ is a package description then we write $\dom(\phi)$ for its domain.
  If $\phi(p,n)=(i,k,d,c,p)$ then we also write
  \begin{itemize}
  \item $\ValInstalled{\phi(p,n)}=i$
  \item $\ValKeep{\phi(p,n)}=k$
  \item $\ValDepends{\phi(p,n)}=d$
  \item $\ValConflicts{\phi(p,n)}=c$
  \item $\ValProvides{\phi(p,n)}=p$
  \end{itemize}
\end{definition}

It is natural to define a package description as a function since we
can have at most one package description for a
given pair of package name and version in a CUDF document. The function is generally
only partial since we clearly do not require to have a package
description for any possible pair of package name and version.

We define the removal operation of a particular versioned package
from a package description. This operation will be needed later in
Definition~\ref{def:consistent} to define the semantics of
\emph{package conflicts} in case a package conflicts with itself or a
feature provided by the same package.

\begin{definition}[Package removal]
  Let $\phi$ be a package description, $p\in\DomPackages$ and
  $n\in\DomVersions$. The package description $\phi-(p,n)$ is defined 
  by
  \begin{eqnarray*}
    \dom(\phi-(p,n)) & = & \dom(\phi) - \{(p,n)\} \\
    (\phi-(p,n))(q,m) & = & \phi(q,m)\qquad \mbox{for all }(q,m)\in \dom(\phi-(p,n))
  \end{eqnarray*}
\end{definition}

\begin{definition}[Request]
  \label{def:request}
  A \emph{request} is a triple $(l_i,l_u,l_d)$ with
  $l_i,l_u,l_d\in\DomCList$.
\end{definition}
In a triple $(l_i,l_u,l_d)$, $l_i$ is the list of packages to be
installed, $l_u$ the list of packages to be updated, and $l_d$ the list
of packages to be deleted.

\subsection{Installations}

\begin{definition}[Installation]
  \label{def:installation}
  An \emph{installation} is a function from \DomPackages{} to
  $P(\DomVersions)$.
\end{definition}

The idea behind this definition is that the function describing an
installation associates the set of
versions that are installed to any possible package name. This set is empty when no version of the
package is installed.

We can extract an installation from any package description as follows:
\begin{definition}[Current installation]
  Let $\phi$ be a package description, the \emph{current package
    installation of $\phi$}
  \[
  i_\phi\colon\DomPackages \rightarrow P(\DomVersions)
  \]
  is defined by
  \[
  i_\phi(p) := \{ n\in\DomVersions \mid (p,n)\in\dom(\phi) \mbox{ and }
  \ValInstalled{\phi(p,n)}=\texttt{true} \}
  \]
\end{definition}

A package can declare zero or more \emph{features} that it provides.
We can also extract the \emph{features} provided by a package
description:
\begin{definition}[Current features]
  Let $\phi$ be a package description, the \emph{current features of $\phi$}
  \[
  f_\phi\colon\DomPackages \rightarrow P(\DomVersions)
  \]
  is defined by
  \[
  f_\phi(p) := \bigcup_{p\in\dom(i_\phi)}\quad(\bigcup_{n\in i_\phi(p)}
  \textrm{expand-features}(\ValProvides{\phi(p,n)})~)
  \]
  where we define
  \begin{eqnarray*}
    \textrm{expand-features}((=,n),f) & = & \{ (f,n) \}\\
    \textrm{expand-features}(\top,f) & = & \{ (f,n) \mid n\in\DomVersions \}
  \end{eqnarray*}
\end{definition}
The second clause in the definition of \textrm{expand-features}
mentioned above expresses the fact that providing a feature without a
version number means providing that feature at any possible version.

In order to define the semantics of a CUDF document, we will frequently
need to merge two installations. This will mainly be used for merging
an installation of packages with an installation of provided features.
The merging operation is formalized as follows:
\begin{definition}[Merging]
  Let $f,g\colon\DomPackages \rightarrow P(\DomVersions)$ be two
  installations. Their merge $f\cup g\colon\DomPackages\rightarrow
  P(\DomVersions)$ is defined as
  \[
  (f\cup g)(p) = f(p) \cup f(p) \qquad\mbox{for any }p\in\DomPackages
  \]
\end{definition}

\subsection{Consistent package descriptions}
We define what it means for an installation to satisfy a constraint:

\begin{definition}[Constraint satisfaction]
  \label{def:satisfaction:constraint}
  The \emph{satisfaction} relation between a natural number $n$ and a
  constraint $c\in\DomConstraints$, noted $n\models c$, is defined as
  follows:
  \begin{center}
  \begin{tabular}{ll@{\qquad\qquad}ll}
    $n\models \top$     & for any $n$
    & $n\models (<,v)$    & iff $n< v$ \\
    $n\models (=,v)$    & iff $n=v$
    & $n\models (>,v)$    & iff $n> v$ \\
    $n\models (\neq,v)$ & iff $n\neq v$
    &  $n\models (\leq,v)$ & iff $n\leq v$ \\
    &&  $n\models (\geq,v)$ & iff $n\geq v$
  \end{tabular}
  \end{center}
\end{definition}

Now we can define what it implies for a package installation to satisfy some
formula:
\begin{definition}[Formula satisfaction]
  \label{def:satisfaction:formula}
  The \emph{satisfaction} relation between an installation $I$ and a
  formula $p$, noted $I\models p$, is defined by induction on the
  structure of $p$:
  \begin{itemize}
  \item $I\models (c,p)$ where, $c\in\DomConstraints$ and
    $p\in\DomPackages$, iff there exists an $n\in I(p)$ such that
    $n\models c$.
  \item $I\models \phi_1\wedge\ldots\wedge\phi_n$ iff $I\models\phi_i$
    for all $1\leq i \leq n$.
  \item $I\models \phi_1\vee\ldots\vee\phi_n$ iff there is an $i$ with
    $1\leq i \leq n$ and $I\models \phi_i$.
  \end{itemize}
\end{definition}

We can now lift the satisfaction relation to sets of packages:
\begin{definition}
  \label{def:satisfaction:lists}
  Let $I$ be an installation, and $l\in \DomCList$. Then $I\models l$
  if for any $(c,p)\in l$ there exists $n\in I(p)$ with $n\models c$.
\end{definition}
Note that, given that $\DomEList\subseteq\DomCList$, this also defines the
satisfaction relation for elements of $\DomEList$. Also note that one
could transform any $l\in\DomCList$ into a formula
$l_\wedge\in\DomForm$, by constructing the conjunction of all the
elements of $l$. The semantics of $l$ is the same as the semantics
of the formula $l_\wedge$.

\begin{definition}[Disjointness]
  \label{def:disjointness:conflicts}
  The \emph{disjointness} relation between an installation $I$ and a
  set $l\in\DomCList$ of packages possibly with version constraints,
  is defined as: $I\parallel l$ if for any $(c,p)\in l$ and all $n\in
  I(p)$ we have that $n\not\models c$.
\end{definition}

\begin{definition}
  \label{def:consistent}
  A package description $\phi$ is \emph{consistent} if for every
  package $p\in\DomPackages$ and $n\in i_\phi(p)$ we have that
  \begin{enumerate}
  \item $i_\phi\cup f_\phi\models \ValDepends{\phi(p,n)}$
  \item $i_{\phi-(p,n)} \cup f_{\phi-(p,n)}\parallel \ValConflicts{\phi(p,n)}$
  \end{enumerate}
\end{definition}

In the above definition, the first clause corresponds to the
\emph{Abundance} property of \cite{edos2006ase}: all the dependency
relations of all installed packages must be satisfied. The second
clause corresponds to the \emph{Peace} property of \cite{edos2006ase}.
In addition, we now have to take special care of packages that
conflict with themselves, or that provide a feature and at the same
time conflict with that feature: we only require that there be no
conflict with any \emph{other} installed package and with any feature
provided by some \emph{other} package (see also
Section~\ref{sec:semantics:comments}).

\subsection{Semantics of requests}

The semantics of a request is defined as a relation between package
descriptions. The idea is that two package descriptions $\phi_1$ and
$\phi_2$ are in the relation defined by the request $r$ if there
exists a transformation from $\phi_1$ to $\phi_2$ that satisfies $r$.
Integration of optimization criteria is discussed in
Section~\ref{sec:cudf-criteria} and is ouside the scope of the current
section.

First we define the notion of a successor of a package description:
\begin{definition}[Successor relation]
  A package description $\phi_2$ is called a \emph{successor} of a
  package description $\phi_1$, noted $\phi_1\rightarrowtail\phi_2$, if
  \begin{enumerate}
  \item $\dom(\phi_1)=\dom(\phi_2)$
  \item For all $p\in\DomPackages$ and $n\in\DomVersions$: if
    $\phi_1(p,n)=(i_1,k_1,d_1,c_1,p_1)$ and
    $\phi_2(p,n)=(i_2,k_2,d_2,c_2,p_2)$ then $k_1=k_2$, $d_1=d_2$,
    $c_1=c_2$, and $p_1=p_2$.
  \item For all $p\in\DomPackages$ 
    \begin{itemize}
    \item for all $n\in i_{\phi_1}(p)$: if
      $\ValKeep{\phi_1(p,n)}=\texttt{'version}$ then
      $n\in i_{\phi_2}(p)$.
    \item if there is an $n\in i_{\phi_1}(p)$ with
      $\ValKeep{\phi_1(p,n)}=\texttt{'package}$ then 
      $i_{\phi_2}(p)\neq\emptyset$
    \item for all $n\in i_{\phi_1}(p)$:
      if $\ValKeep{\phi_1(p,n)}=\texttt{'feature}$ then 
      $i_{\phi_2}\cup f_{\phi_2}\models \ValProvides{\phi_1(p,n)}$
    \end{itemize}
  \end{enumerate}
\end{definition}
The first and the second item of the above definitions indicate that a
successor of a package description $\phi$ may differ from $\phi$ only
in the status of packages. The third item refines this even further
depending on keep values:
\begin{itemize}
\item If we have a keep status of \texttt{version} for an installed
  package $p$ and version $n$ then we have to keep that package and version.
\item If we have a keep status of \texttt{package} for some installed version
  of a package $p$ then the successor must have at least one version of that
  package installed.
\item If we have a keep status of \texttt{feature} for some installed
  version $n$ of a package $p$ then the successor must provide all the
  features that where provided by version $n$ of package $p$.
\end{itemize}

\begin{definition}[Request semantics]
  Let $r=(l_i,l_u,l_d)$ be a request. The \emph{semantics} of $r$ is a
  relation $\ReqSem{r} \subseteq \PackageDescriptions \times
  \PackageDescriptions$ defined by $\phi_1 \ReqSem{r} \phi_2$ if
  \begin{enumerate}
  \item $\phi_1 \rightarrowtail\phi_2$
  \item $\phi_2$ is consistent
  \item $i_{\phi_2}\cup f_{\phi_2}\models l_i$
  \item $i_{\phi_2}\cup f_{\phi_2}\parallel l_d$
  \item $i_{\phi_2}\cup f_{\phi_2}\models l_u$, and for all $p\in l_u$
    and all $(p,n)\in l_u$ we have that $i_{\phi_2}(p)=\{n\}$ is a
    singleton set with $n\geq n'$ for all $n'\in i_{\phi_1}(p)$.
  \end{enumerate}
\end{definition}

\subsection{Comments on the semantics \NONNORM}
\label{sec:semantics:comments}

\paragraph{Installing multiple versions of the same package}
The semantics allows a priori to install multiple versions of the same
package. This coincides with the semantics found in RPM-like F/OSS
distributions (which a priori do not forbid to install multiple
versions of the same package), but is in opposition to the semantics
found in Debian-like F/OSS distributions (which allow for one
version of any package to be installed at most).

In many practical cases the distinction between a priori allowing or
not for multiple versions of a package makes little difference. In
the RPM world multiple versions of the same package are very often in
a conflict by their features or shipped files. If both versions of the
same package provide the same feature and also conflict with that
feature then the RPM semantics, as the CUDF semantics, does not allow
to install both at the same time. Only packages that have been
designed to have distinct versions provide distinct features (in
particular, files with distinct paths) can in practice be installed in
the RPM world in several different versions at a time. This typically
applies to operating system packages. In order to have a
meta-installer with Debian semantics work correctly on such a package
description, it is sufficient to rename the packages, and to create a
new package, say $p-n$, for a package $p$ and version $n$ when $p$ can
be installed in several versions.

On the other hand, a meta-installer with RPM semantics will produce solutions on
a package description that would not be found by a
meta-installer with Debian semantics since it is free to install
several version of the same package. The uniqueness restriction of
Debian can easily be made explicit in the package description by
adding say a serialized property ``\texttt{Conflicts:} $p$''to any stanza in the package description, say of package $p$,.

\section{Integrating optimization criteria}
\label{sec:cudf-criteria}

The semantics given in Section~\ref{sec:cudf-semantics} is designed to
define when passing from one installation to another installation
satisfies a user request. It does not discriminate among different
resulting installations, which is in most cases too coarse to express
the requirements a good meta-installer should satisfy. For instance,
one might expect from a meta-installer that it does not call for the
installation of unnecessary packages, or that it installs the latest
version of packages when possible.

All single criteria of this kind can easily be expressed as
optimization criteria that are in fact orthogonal to the semantics
defined in Section~\ref{sec:cudf-semantics}. In order to express
optimization criteria we use an optional property of package
descriptions, called \texttt{Cost}, for instance, with value space \texttt{int}
and default $0$. Using this new property we can extend the definition
of a package description to also yield the value of the \texttt{Cost}
property in addition to the five properties already required in
Definition~\ref{def:package:description}. We will write
$\ValCost{\phi(p,v)}$ for the value of the property \texttt{Cost} of
the package description with package name $p$ and version $n$.

\begin{definition}[Installation cost]
  The \emph{cost} $\textit{cost}(\phi)$ of a package description
  $\phi$  is defined as
  \[
  \textit{cost}(\phi) = \sum_{p\in\DomPackages}\quad(\sum_{v\in i_\phi(p)}
  \ValCost{\phi(p,v)}~)
  \]
  We say that $\phi_1$ is \emph{as most as expensive as} $\phi_2$,
  written $\phi_1 \lesssim \phi_2$, if $\textit{cost}(\phi_1) \leq
  \textit{cost}(\phi_2)$.
\end{definition}
In other words, the cost of a package description is simply the sum of
the \texttt{Cost} values of all installed versions of
packages. Mathematically, the relation $\lesssim$ is a so-called
quasi-order, that is we have that $\lesssim$ is reflexive ($\phi
\lesssim \phi$ for all $\phi$) and transitive ($\phi_1 \lesssim
\phi_2$ and $\phi_2 \lesssim \phi_3$ imply $\phi_1 \lesssim \phi_3$),
but not necessarily anti-symmetric (it may be the case that $\phi
\lesssim \psi$ and also $\psi \lesssim \phi$ for different package
descriptions $\phi$ and $\psi$).

\begin{lemma}
  \label{lem:optimization}
  Let $\phi$ be a package description such that $\dom(\phi)$ is
  finite, and $r$ a request. If there exists an $\psi$ such that $\phi
  \ReqSem{r} \psi$ then there exists an installation $\psi_0$ such
  that $\psi_0 \lesssim \psi$ for all $\psi$ with $\phi \ReqSem{r}
  \psi$.
\end{lemma}
This means that if a request has a solution at all then
an optimal solution exists, even though this optimal solution may not be
unique.

The proof of Lemma~\ref{lem:optimization} is obvious from the fact
that the smallest possible value of $\textit{cost}(\psi)$ is limited
by
\[
\sum_{(p,n)\in\dom(\phi)} \min\{0,\ValCost{\phi(p,n)}\}
\]
Alternatively, one might argue that for any given finite package
installation there is only a finite number of possible successors.

Let us now see how various frequent optimization criteria can be
translated into appropriate choices of the \texttt{Cost} property.

\begin{itemize}
\item
  Optimization of the disk space occupied by the installation may be
  indicated by putting the size taken by the installation of a
  package as value of the
  \texttt{Cost} property in the package description.\index{}
\item
  Optimization of the download size required to pass to the new
  installation may be indicated by putting the download size of the package as the value of the
  \texttt{Cost} property if that
  version of the package is not installed, and $0$ if that version of
  the package is already installed. Note that this amounts to
  expressing an optimization criterion for all \emph{newly} installed
  packages even though the general optimization mechanism is defined
  on the set of \emph{all} packages that are installed in the end.
\item 
  Preference of installation of most recent available versions of packages
  can be expressed by putting the value $0$ as the value of the \texttt{Cost}
  property of package $p$ and version $v$ if $n$ is the
  latest version of $p$, and $1$ otherwise. This can easily be
  extended to taking into account ``how outdated'' a package is,
  either by putting the value of \texttt{Cost} of package~$p$ and
  version~$v$ to be the number of versions of~$p$ that are greater or
  equal to~$v$ and strictly smaller than the latest version of~$p$, or
  by using some other metric.
\item
  The requirement of installing a minimal number of auxiliary packages
  (i.e. packages that are not mentioned in the request) can be
  implemented by putting the value of \texttt{Cost} property to $0$ if
  version $v$ of package $p$ is already installed or if its
  installation is explicitly required, and $1$ otherwise.
\item
  The requirement that a minimal number of packages should be removed is
  implemented by putting the value of the \texttt{Cost} property to $-1$
  if a package is installed, and to $0$ otherwise.
\end{itemize}
Note that all these optimization criteria are \emph{single} criteria.
A user might have a vague notion of wanting to optimize several of
these criteria at the same time (such as ``upgrade as many packages as
possible to their latest version, and at the same time remove as few
packages as possible''). However, it is at the moment absolutely not
clear what the exact semantics of this might be. In order to define
the semantics of such a request one would have to define, for any two
solutions, which of the two is the preferred solution.

Also note that while we have used a single \texttt{Cost} property
throughout this section to discuss optimization possibilities,
implementations are not required to do so. In fact, it is recommended
that implementations define meaningful properties (e.g.
\texttt{Installed-Size}, \texttt{Download-Size}, etc.) and that cost
functions to be optimized get defined over the semantic values assumed
by those extra properties.

\section{Serialization}
\label{sec:cudf-serialization}

This section describes how to serialize CUDF documents as stream of
bytes and, symmetrically, how to parse streams of bytes as CUDF
documents. We refer generically to one or the other action as
\emph{CUDF serialization}.

Serialization is meant to make the storage of CUDF documents as
files possible and to transfer them over the network. A stream of bytes which
can be parsed as a CUDF document respecting this specification is
called a \emph{CUDF file}.

\subsection{Overall CUDF file syntax}

A CUDF file is a plain-text file containing several \emph{file
  stanzas}. The bytes composing the file should be interpreted as
Unicode characters encoded in UTF-8.

The overall organization of a CUDF file in stanzas reflects the
schematic structure of CUDF content (see
Section~\ref{sec:cudf-infoitems}). Each file stanza is the
serialization of a CUDF information item. Blank lines (i.e. empty
lines, or lines composed only by white space characters:
\texttt{U+0020}, \texttt{U+000D}, \texttt{U+000A}, \texttt{U+0009})
occurring between file stanzas are ignored.

Serialization should---where possible---follow the canonical ordering of
information items given in Section~\ref{sec:cudf-infoitems}, that is
first contain the list of stanzas corresponding to package
descriptions (\emph{package description stanzas}) and then the sole
stanza corresponding to problem description (\emph{problem description
  stanza}).

To recognize the beginning of file stanzas, each of them starts with a
\emph{postmark}, which is specific to information item
classes. Postmarks denote the beginning of a new file stanza only
\emph{when occurring either at the beginning of the file or just after
  a newline} (Unicode code point \texttt{U+000A}).
\begin{itemize}
\item For package description items, the postmark is the string
  \LEX{Package: }.
\item For problem description items, the postmark is the string
  \LEX{Problem: }.
\end{itemize}
In both cases, the postmark can be followed by some characters other
than a newline, and end with a single newline.

\subsection{Information item serialization}

Each information item, whatever its class, is serialized as a stream
of bytes serializing all of its properties in an arbitrary order. A
single property is serialized as a stream of bytes performing the
following steps in order:
\begin{enumerate}
\item serialize the property name as the string corresponding to the
  \textsc{Name} given in its property schema;
\item output the string \LEX{: }, i.e. a double colon followed by a
  space (\texttt{U+0020});
\item serialize the property value;
\item output a single newline.
\end{enumerate}

Let $t$ be the type of a property whose value $v$ has to be serialized
as a stream of bytes. The value is serialized by choosing a value from
$v'\in\mathcal{L}(t)$ such that $\mathit{parse}_t(v')=v$. That is, all
possible values that will be parsed back as the value to be serialized
are valid serializations of it.

Since parsing is not one-to-one in general for CUDF types, it is
possible that different implementations of this specifications make
different choices in terms of how to serialize a given value. Hence it
should not be taken for granted that two serializations of CUDF values
which are not byte-to-byte identical do not denote the same CUDF
value.

An important distinction exists between the serialization of different
classes of information items. For package descriptions, the postmark
is part of the serialization of properties, i.e. the line starting
with \LEX{Package: } is the serialization of the \texttt{Package}
property (i.e. the package name). As a consequence, and in amendment
of the general rule above on the property serialization order, the
\emph{\texttt{Package} must be the first property serialized in each
  file stanza}, because it is used to recognize the beginning of
package description file stanzas.

On the contrary, for problem descriptions the postmark is used to
recognize the beginning of the corresponding file stanza, but does not
represent the serialization of any particular property. Instead of
leaving an empty line after the problem description postmark,
implementations should output a problem identifier, possibly
cross-referencing the source from which a given CUDF is being
generated from (e.g. a DUDF unique identifier).

An example of CUDF file is given in
Appendix~\ref{chap:cudf-serialization}.

\subsection{CUDF file parsing}

How to parse CUDF files to obtain CUDF documents is almost
straightforward and follows from an analysis of the serialization
rules given above.

\emph{Parsing errors} can be encountered while parsing CUDF
serializations which do not match the rules provided by this
specification. Parsing errors can be localized at specific positions
of the CUDF serialization. When the position of a parsing error
belongs to a specific file stanza (i.e. it is in between two
postmarks, or between a postmark and the end of file), that error is
said to be recoverable. The recovery strategy is to ignore the file
stanza the error belongs to and act as if that stanza was not there.

The actual parsing procedure is as follows:
\begin{enumerate}
\item Given a CUDF file, split it at occurrences of postmarks. The
  result of this operation is a list of file stanzas. Each of them can
  be recognized as the serialization of either a package description
  (if the postmark is \LEX{Package: }) or a problem description (if
  the postmark is \LEX{Problem: }).

  Afterwards, problem description postmarks are useless and can be
  ignored for further processing. On the contrary, problem description
  postmarks should be integrated again as part of the following
  package description file stanza.

\item Parse each file stanza as a list of property serializations by
  splitting at occurrences of newlines.

\item Parse each property serialization as a pair of property name and
  value serializations by splitting when the string \LEX{: } occurs.

\item Turn each property name serialization into a property name in a
  straightforward way, as long as it matches the constraints on
  property names. Otherwise raise a parsing error; the error must be
  located in the file stanza owning the affected property.

  For each property name check whether that property is supported by
  the information item serialization in which it appears. If this is
  the case then this specification permits to assign all the
  information coming from its schema to that property, in particular a
  type and possibly an optional value.  If the property is not
  supported by this specification for a given information item, it is
  either known, via some unspecified external mechanism, how to
  associate a schema to that property or that property cannot be
  processed any further and will be disregarded.

  After this step all properties have an associated schema and a (yet
  to be parsed) value serialization.

\item For each value serialization parse it using the parsing function
  of the associated property type. If the function is not defined for
  the given serialization then a parsing error is raised; the error is
  located in the file stanza owning the affected value.

  After this step, each file stanza has been parsed into a list of
  properties as supported by CUDF. That list can be turned into a set.
  If the same property name appears twice or more in the set, a parsing
  error is raised; the error is located in the file stanza containing
  the properties.

  Once sets are formed, the CUDF file has been fully parsed into a
  list of information items; together they already form a CUDF
  document.

\item The only missing step is handling of default values. For each
  information item check whether some of the optional properties are
  missing according to the information item kind (package or problem
  description). For each such missing optional property, add a
  property of that name to the information item where it was
  missing. The corresponding value is either \texttt{None} (for
  properties which do not specify a default value) or the default
  value defined in the property schema.

\end{enumerate}

\subsection*{Compatibility with RFC822 \NONNORM}

Conforming implementation of CUDF serialization produces CUDF files
which are blank-separated sequences of messages conforming to
RFC822~\cite{rfc822}.

This aspect hints an alternative---yet correct---way of parsing CUDF
files via exploitation of existing RFC822 implementations. On top of
them it is enough to perform the parsing steps given above from 4 to
6, skipping steps 1--3 which are subsumed by RFC822 parsing.

\section{Generating CUDF \NONNORM}
\label{sec:cudf-conversion}

While it is possible to generate CUDF documents directly, it is
expected that the largest fraction of the CUDF corpus to be used for
the competition will be generated via conversion from DUDF documents
provided by users of F/OSS distributions.

Each distribution interested in providing upgrade problems for the
UPDB (see Chapter~\ref{chap:intro}) is then required to provide
specification and tools that implement the conversion. Ideally, the
description of how to convert from a specific DUDF instance and CUDF
should be described together with the specification of the specific
DUDF instance. It is expected that each partner interested in
contributing problems to the UPDB publishes a document describing both
aspects.

During the conversion, we expect three main tasks to be implemented.
\begin{description}
\item[Translation: package information $\longrightarrow$ package
  information items] Each DUDF instance is expected to encode the
  information about all packages known to the meta-installer in some
  way. The first required task to create the resulting CUDF is to
  convert such (meta-installer-/distribution-specific) information to
  package information items as described in this specification.

  The implementation of this task should account not only for data
  conversions imposed by the CUDF type system (e.g. translating from
  legacy versioning schemata---\emph{x}.\emph{y}.\emph{z}---to
  positive integers), but also for semantic differences between the
  origin distribution and CUDF. Likely, the most common cause of
  semantic incompatibilities will be the translation from Debian-like
  packaging systems to CUDF (see Section~\ref{sec:semantics:comments}
  for advice on how to address this problem).

\item[Translation: user request $\longrightarrow$ problem description
  item] Similarly, the request that the user posed to its
  meta-installer needs to be translated to a problem description
  item.

  The request language supported by CUDF is expected to be expressive
  enough to encode the vast majority of user requests nowadays
  expressible in state of the art meta-installers. Exceptions are of
  course possible, in which case no translation from DUDF to CUDF is
  possible. Specifications of DUDF instances must clearly state such
  limitations.

\item[Expansion of intentional sections] DUDF encodings are expected
  to be more compact than the corresponding CUDF encoding (see
  Section~\ref{sec:dudf-intension}). To that end DUDF documents can
  refer to external resources whereas CUDF documents are expected to
  be entirely self-contained. Therefore, all references to external
  entities occurring in DUDF documents must be expanded before being
  able to create the corresponding CUDF encoding.

  Since in general only distributions are expected to be able to
  perform the expansions (e.g. because the referred repositories are
  mirrors or databases hosted by them), the actual translation from
  DUDF instances to CUDF should be performed by distributions
  \emph{before} injecting problems into the central UPDB.

\item[Serialization] Once all information items translated from DUDF
  to the CUDF model, they need to be serialized to files (see
  Section~\ref{sec:cudf-serialization}).

\end{description}



\chapter{Conclusion}
\label{chap:outro}

The \mancoosi{} project will run a solver
competition~\cite{mancoosi-wp1d1}, in which each participant will try
to find the best possible solutions to upgrade problems as those faced
by users of F/OSS software distributions.  This specification has
defined two (classes of) document formats which play important roles in
the work-flow of the competition.

The first class of document formats is DUDF (Distribution Upgrade
Description Format), described in
Chapter~\ref{chap:dudf}. Specific instances of DUDF will be used as
document formats to encode real life problems encountered by users of
F/OSS software distributions. DUDF is meant to be a compact
representation of upgrade problems, suitable to be transferred over
the network. In addition to the purposes of the competition, DUDF documents
might be useful to store and transfer the state of package managers,
for example for reporting bugs concerning package managment tools.

Distributions that are interested in providing problems on which the
competition will possibly be run should have an interest in implementing
DUDF for their own distributions. The current document only
describes the outline (or skeleton) of DUDF. Implementing DUDF
actually means standardizing a specific instance of it, by describing
in a separate document how the holes left open by this specification
have to be filled in the context of a specific software
distribution. Equipped with this specification and the specification
of a DUDF instance, implementors will be able to produce and interpret
DUDF corresponding to upgrade problems faced on final user machines.

The second document format introduced by this specification if CUDF
(Common Upgrade Description Format), described in
Chapter~\ref{chap:cudf}. The purpose of CUDF is to provide a model in
which upgrade problems can be encoded, by abstracting over
distribution-specific details. In the context of the competition, the
interest of CUDF is to encode problems on which the actual competition
will be run. This way, participating solvers will not need to implement
distribution-specific semantics, and will only have to reason about a
self-constained problem.

As far as CUDF is concerned, this specification has provided a formal model in which
constraints coming from popular packaging ``worlds'' (e.g. Debian and
RPM) can be expressed. On top of that model the semantics of typical
upgrade action requests (e.g. install, remove, upgrade) has been
described; using that semantics it is possible to check whether a
solution provided by a solver properly implements a given user
request.

In addition to the formal model, this specification has also provided
a document structure in which both the user request and the universe
of all packages known to a package manager can be encoded. Parsing and
serialization rules for the document structure have been given as well,
so that CUDF documents can be dealt with in file form. Solvers taking
part in the competition are meant to parse CUDF files in order to
obtain the upgrade problem they are asked to solve.

To complete the competition scenario two important aspects are
missing, but have been left beyond the scope of this document on purpose:
\begin{description}

\item[Optimization criteria] It is expected that solvers taking part
  into the competition will not simply be asked to solve a given
  upgrade problem. At least for some competition ``tracks'', there
  will be extra requirements to find the best possible solution among
  several alternative solutions which are correct according to the
  CUDF model.

  How to specify optimization criteria is beyond the scope of this
  document and, is also outside the purpose of CUDF files. Each
  competition track will advertise the optimization criteria to be
  implemented by participating solvers. Optimization criteria can be
  defined on top of package properties which are already expressible
  in the present version of CUDF. To this end, CUDF is extensible:
  additional properties not prescribed by this specification can be
  added to package descriptions, by exploiting existing CUDF types.

\item[Solver output format] The output format of solvers taking part
  in the competition is beyond the scope of this
  specification. Nevertheless it will be needed in order to have a
  common way to understand the solutions found by solvers and to
  determine their quality according to the optimization criteria.

  Naively, the solver output can be encoded by serializing the new
  local package status as if it were a CUDF document missing the
  problem description item. Practically though, such a representation
  would encode a lot of information which is a duplication of the CUDF
  input initially fed into the solver. Hence, a format which is more
  likely to be used for solver output is a ``patch'' with respect to
  the initial local package status as encoded in the CUDF input.

  A separate document will be published, well in time for the
  competition, to describe the required output format and how to
  interpret it to obtain the package status meant by the solver.

\end{description}

\appendix
\chapter{DUDF skeleton serialization example}
\label{chap:dudf-serialization}

\begin{figure}[t!]
\begin{lstlisting}
<dudf version="1.0"
    xmlns="http://www.mancoosi.org/2008/cudf/dudf"
    xmlns:dudf="http://www.mancoosi.org/2008/cudf/dudf">
  <timestamp><!-- timestamp in RFC822 format --></timestamp>
  <uid><!-- unique problem identifier --></uid>
  <distribution><!-- distribution identifier --></distribution>
  <installer>
    <name><!-- installer name --></name>
    <version><!-- installer version --></version>
  </installer>
  <meta-installer>
    <name><!-- meta-installer name --></name>
    <version><!-- meta-installer version --></version>
  </meta-installer>
  <problem>
    <package-status>
      <installer><!-- installer status --></installer>
      <meta-installer>
        <!-- meta-installer status -->
      </meta-installer>
    </package-status>
    <package-universe>
      <package-list
          dudf:format=<!-- package list format identifier -->
          dudf:filename=<!-- package list absolute path --> >
        <!-- package list -->
      </package-list>
      <!-- ... other package lists ... -->
      <package-list
          dudf:format=<!-- package list format identifier -->
          dudf:filename=<!-- package list absolute path --> >
        <!-- package list -->
      </package-list>
    </package-universe>
    <action><!-- requested meta-installer action --></action>
    <desiderata><!-- meta-installer desiderata --></desiderata>
  </problem>
  <outcome dudf:result=<!-- one of: "success", "failure"--> >
    <error><!-- error description (result: "failure")--></error>
    <package-status> <!-- result: "success" -->
      <installer><!-- new installer status --></installer>
      <meta-installer>
        <!-- new meta-installer status -->
      </meta-installer>
    </package-status>
  </outcome>
</dudf>
\end{lstlisting}
  \caption[XML serialization of DUDF]{XML serialization skeleton of a
    DUDF problem/outcome submission}
  \label{fig:dudf-xml}
\end{figure}

This non-normative section contains an example of DUDF serialization
to XML. The example is given in Figure~\ref{fig:dudf-xml}, which is
the serialization of the DUDF skeleton given in
Figure~\ref{fig:dudf-xml}. In the example, XML comments have been put
in place of outline holes and other missing information.

\chapter{RELAX NG schema for DUDF}
\label{chap:dudf-relax}

This non-normative section contains a RELAX NG~\cite{relaxng} schema
which can be used to check whether a given XML document represents a
valid DUDF skeleton serialization. The schema only ensures that the
skeleton part of the XML document is valid with respect to this
specification, since the details about how holes are filled are
distribution-specific.

Additional comments in the schema denote ``side
conditions''---e.g. the fact that dates should be in RFC882
format---which are not expressed by the schema itself, and which should be
checked to ensure proper implementation of DUDF.

\begin{figure}[t!]
\begin{lstlisting}
default namespace dudf = "http://www.mancoosi.org/2008/cudf/dudf"

any = ( element * { any* } | attribute * { text }* | text )

tool_id = (
   element name { text },	# must be a package name
   element version { text }	# must be a version number
)

package_status =
   element package-status {
      element installer { any* },	# installer-specific
      element meta-installer { any* }?	# meta-installer-specific
   }

start = element dudf {
   attribute dudf:version { "1.0" },
   element timestamp { text },	# must be a date in RFC822 format
   element uid { text },
   element distribution { text },
   element installer { tool_id },
   element meta-installer { tool_id },
   element problem {
      package_status,
      element package-universe {
         element package-list {
            attribute dudf:format { text },
            attribute dudf:filename { text }?,	# must be an
            					# absolute path
            any*
         }+
      },
      element action { text },
      element desiderata { text }?
   },
   (element outcome {
      attribute dudf:result { "success" },
      package_status
   }
    | element outcome {
       attribute dudf:result { "failure" },
       element error { text }
    })
}
\end{lstlisting}
  \caption[RELAX NG schema for DUDF]{RELAX NG schema for the DUDF
    skeleton}
  \label{fig:dudf-relax}
\end{figure}

The RELAX NG schema is reported in Figure~\ref{fig:dudf-relax}.

\chapter{CUDF serialization example}
\label{chap:cudf-serialization}

This non-normative section contains an example of CUDF serialization
to file. The example below has been inspired by the
EDOS car/glass example~\cite{edos-wp2d2}.

Some remarks about the example follow.
\begin{itemize}
\item The example does not rely on any extended properties.
\item Intuitively, the example comes from a packaging world where
  different versions of the same package are implicitly conflicting
  with each other. To grasp this, all packages for which multiple
  versions are available declare a non-versioned conflicts with
  themselves.
\item The \texttt{engine} feature is mutually exclusive, only one
  (installed) package can provide it. This is encoded using conflicts
  with the feature from each package providing it.
\end{itemize}

\begin{lstlisting}
Package: car
Version: 1
Depends: engine, wheel, door, battery
Installed: true

Package: bicycle
Version: 7

Package: gasoline-engine
Version: 1
Depends: turbo
Provides: engine
Conflicts: engine, gasoline-engine
Installed: true

Package: gasoline-engine
Version: 2
Provides: engine
Conflicts: engine, gasoline-engine

Package: electric-engine
Version: 1
Depends: solar-collector | huge-battery
Provides: engine
Conflicts: engine, electric-engine

Package: electric-engine
Version: 2
Depends: solar-collector | huge-battery
Provides: engine
Conflicts: engine, electric-engine

Package: solar-collector
Version: 1

Package: battery
Version: 3
Provides: huge-battery
Installed: true

Package: wheel
Version: 2
Conflicts: wheel
Installed: true

Package: wheel
Version: 3
Conflicts: wheel

Package: door
Version: 1
Conflicts: door
Installed: true

Package: door
Version: 2
Depends: window
Conflicts: door

Package: turbo
Version: 1
Installed: true

Package: tire
Version: 1
Conflicts: tire

Package: tire
Version: 2
Conflicts: tire

Package: window
Version: 1
Conflicts: window

Package: window
Version: 2
Depends: glass = 1
Conflicts: window

Package: window
Version: 3
Depends: glass = 2
Conflicts: window

Package: glass
Version: 1
Conflicts: glass

Package: glass
Version: 2
Conflicts: glass, tire = 2

Problem: source: Debian/DUDF 733963bab9fe1f78fd551ad20485b217
Install: bicycle, electric-engine = 1
Upgrade: door, wheel > 2
\end{lstlisting}

\bibliography{mancoosi}
\bibliographystyle{alpha}

\end{document}